\documentclass[useAMS,usenatbib]{mnras}
\usepackage{newtxtext,newtxmath}
\usepackage{ae,aecompl}
\usepackage{graphicx,epsfig,psfig}
\usepackage[normalem]{ulem}
\usepackage{xcolor,verbatim}
\usepackage{amsmath}	
\usepackage{tabularx}

\def \tt {\textit}
\def \cstat {\textit{cstat} }
\def \xspec {\textsc{xspec} }

\newcommand{\comments}[1]{}

\newcommand {\pass}{\textsc{pass} }

\def\msun {\rm M_\odot}

\newcommand {\ergps} {erg s$^{-1}$ }
\newcommand {\kmps} {km s$^{-1}$ }

\newcommand {\tctf} {$t_{\rm cool}/t_{\rm ff}$ }
\newcommand {\tc} {$t_{\rm cool}$ }
\newcommand {\tf} {$t_{\rm ff}$ }
\newcommand {\mtctf} {min$(t_{\rm cool}/t_{\rm ff})$ }
\newcommand {\nlfit}{\textit{nl-fit} }
\newcommand {\mcfit}{\textit{mcmc-fit} }

\definecolor{webgreen}{rgb}{0,.5,0}
\definecolor{webbrown}{rgb}{.6,0,0}

\defcitealias{lakhchaura16}{L16}
\defcitealias{hogan17}{H17}

\title[Deprojection of X-ray data in galaxy clusters]{Deprojection of X-ray data in galaxy clusters: confronting simulations with observations}  

\author[Sarkar, Dey \& Sharma]
{Kartick C. Sarkar \thanks{\url{sarkar.kartick@mail.huji.ac.il}, \url{kartick.c.sarkar100@gmail.com}}$^1$, Arjun Dey\thanks{\url{arjun.dey@weizmann.ac.il}}$^{2}$, Prateek Sharma\thanks{\url{prateek@iisc.ac.in}}$^{3}$\\
$^1$ Racah Institute of Physics, The Hebrew University of Jerusalem, Israel \\
$^2$ Faculty of Physics, Weizmann Institute of Science, Israel \\
$^3$ Department of Physics and Joint Astronomy Program, Indian Institute of Science, Bangalore , India 560012 }

\begin{document}
\maketitle

\label{firstpage}
\begin{abstract}
Numerical simulations with varying realism indicate an emergent principle -- multiphase condensation and large cavity power occur when the ratio of the cooling time to the free-fall time (\tctf) falls below a threshold value close to 10. Observations indeed show cool-core signatures when this ratio falls below 20-30, but the prevalence of cores with  \tctf~ratio below 10 is rare as compared to simulations. 
In X-ray observations, we obtain projected 
spectra from which we have to infer radial gas density and temperature profiles. Using idealized models of X-ray cavities and multiphase gas in the core and 3-D hydro jet-ICM simulations, we quantify the biases introduced by deprojection based on the assumption of spherical symmetry in determining \tctf. 
We show that while the used methods are able to recover the \tctf ratio for relaxed clusters, they have an uncertainty of a factor of $2-3$ in systems containing large cavities ($\gtrsim 20$ kpc). We also show that the mass estimates from these methods, in the absence of X-ray spectra close to the virial radius, suffer from a degeneracy between the virial mass ($M_{200}$) and the concentration parameter ($c$) in the form of $M_{200} c^2 \approx$ constant. 
Additionally, lack of soft-X-ray ($\lesssim 0.5$ keV) coverage and 
poor spatial resolution make us overestimate min(\tctf) by a factor of few in clusters with min(\tctf) $\lesssim 5$. This bias can largely explain the lack of cool-core clusters with min(\tctf) $\lesssim 5$. 

\end{abstract}


\section{Introduction}

Multi-wavelength observations (e.g., \citealt{peterson2003,Raf2008,cav08,odea2008}) and numerical simulations (e.g., \citealt{sharma12,gaspari12,Li2015,prasad15}) have 
led to a consensus picture of the physics of cool cluster cores. The radio bubbles blown by the active
galactic nucleus (AGN) jets powered by accretion onto the central supermassive black hole (SMBH) carve out cavities in X-rays and inject power in the intracluster 
medium (ICM) comparable to the radiative losses in the core (\citealt{churazov2002,birzan2004}). Thus the cluster core is in rough thermal balance, and the presence of
cold gas can most naturally be understood as a consequence of the nonlinear saturation of local thermal instability in presence of gravity and uplift due to AGN jets. For small density perturbations, 
cold gas condenses out of the hot ICM if the ratio of the background cooling time to the free-fall time (\tctf) is smaller than a threshold close to 10 (\citealt{sharma12,choudhury2016}). For larger \tctf~only 
small density/temperature perturbations and internal gravity waves are produced (\citealt{mccourt12,voit17}). But nonlinear density perturbations and uplift by jets can produce multiphase gas even for larger \tctf (\citealt{mcnamara2016,choudhury2019}).

The idealized simulations with heating and cooling balanced in shells are too idealized to compare with observations in detail but provide a useful physical framework of local 
thermal instability in presence of global thermal balance and background gravity. The biggest shortcomings of this model are the absence of angular momentum and strict thermal balance. Realistic simulations
with AGN jets tied to cooling/accretion rate at the center (dominated by the gas in the cold phase) show departure from a perfect thermal balance in the form of cooling and heating cycles 
(\citealt{prasad15,Li2015}). Also, because of the small angular momentum imparted by jets, the condensing cold gas has non-zero angular momentum and  most of the cold gas 
circularizes at a radius  $\sim$ 1 kpc, much larger than the SMBH sphere of influence. Thus, the question of angular momentum transport and the eventual accretion onto the SMBH, which is crucial for feedback heating,
calls for closer scrutiny (e.g., see \citealt{hobbs2011,gaspari13,prasad17}).

Recent observations (\citealt{hogan17,pulido18}) have very carefully mapped out the gravitating mass distribution in galaxy cluster cores, particularly including the contribution of the
brightest central galaxy (BCG) typically found at the centers of relaxed clusters. Accurate determination of the gravitational acceleration ($g$), and hence $t_{\rm ff} \equiv (2r/g)^{1/2}$, 
profile is crucial in testing the \tctf~models. Earlier models (e.g., \citealt{voit2015,lakhchaura16}) were not as accurate in calculating the $t_{\rm ff}$ profile close to the center (where \tctf~is minimum). 
The key advantage of the \citet{hogan17,pulido18} models is that they fix the central potential to be that of an isothermal sphere commensurate with the stellar mass of the central BCG.
In other methods, where either only the dark matter halo potential is used or where a parametric form of the potential is not assumed, imposing hydrostatic equilibrium (HSE) can underestimate gravitational acceleration.
Additionally, \cite{hogan17} find that the number of observed clusters below \tctf$\lesssim 10$ is significantly lower than what was 
found in the numerical simulations \citep[for example][]{Li2015, prasad18},  implying that the real clusters are not cooling as much as in simulations. 
There could be several possible reasons for such a discrepancy including, but not limited to, some missing physics in the simulations and biases in interpreting the observed spectra. We require detailed 
analysis of these possibilities to reconcile this discrepancy.

In this paper, we explore the effect of observational biases in interpreting non-hydrostatic atmospheres and realistic simulations. We use different density and temperature distributions -- ranging from smooth profiles in hydrostatic equilibrium to very disturbed ICM in jet-ICM simulations -- to quantify 
biases introduced by spectral deprojection and different methods to obtain $t_{\rm ff}$ profiles. We use the mock projected spectra and compare the methods of \citet[][hereafter \citetalias{hogan17}]{hogan17} and \citet[][hereafter, \citetalias{lakhchaura16}]{lakhchaura16} in recovering  the input \tctf profiles. We also use the data from realistic simulations \citep{prasad15} that include AGN feedback to compare the  
true \tctf profiles against the recovered \tctf profiles using the observational/analysis techniques. We also compare methods used by \citetalias{lakhchaura16} and \citetalias{hogan17} to recover the \tctf profiles and point out their strengths and weaknesses in each of the cases. 

The paper is organized as follows. We describe our methods of creating projected spectra from a number of theoretical atmospheres in section \ref{sec:method}. Section \ref{sec:results} describes the results of our experiments in each case and provides an overall understanding of biases in recovering \tc and \tf. We discuss the implications of our results in section \ref{sec:discussion}. We finally, present our conclusions in section \ref{sec:conclusions}.

%
%

%
\section{Method}
\label{sec:method}
In this paper, we wish to generate mock projected X-ray spectra from theoretical/simulated ICM models and to understand the biases introduced by various assumptions.
This is done in four steps. First, projected surface brightness maps and spectra are obtained for a given theoretical/simulated model using a software package called \textsc{pass} \citep{sarkar17} assuming \textsc{apec} plasma emission model. The spectra are then divided into different annuli so that each annulus contains a minimum number of counts. Next, we find the average spectrum for each annulus, which is then fitted and de-projected to obtain a radial profile for the electron density and temperature.  Finally, the cluster gravitational models and other quantities are  reconstructed using these density and temperature profiles. Below we describe these procedures one by one.

\subsection{Models}
\label{subsec:toy-models}
To understand the biases, we consider a variety of toy models as well as realistic simulation data of galaxy clusters. Below we describe each of these models.

\begin{table*}
	\centering
	\label{table:test-models}
	\begin{tabular}{p{2cm} p{8cm} p{5cm}}
		\hline \hline
		Model Name & Description & Values \\ 
		\hline \hline
		sc1a & Spherical cavity with radius $r_0$ and placed at $z = \pm r_0$; min(\tctf) $\approx 12$ & $r_0 = 10$ kpc \\
		sc2a & "	& $r_0 = 20$ kpc \\
		sc3a & "	& $r_0 = 40$ kpc \\
		\hline
		cc1a & Conical cavity with half opening angle $\theta_{\rm cav}$ and height $h_{\rm cav}$;  min(\tctf) $\approx 12$ 	& $h_{\rm cav} = 10$ kpc, $\theta_{\rm cav} = 30^\circ$\\
		cc2a & "	& $h_{\rm cav} = 20$ kpc, $\theta_{\rm cav} = 30^\circ$\\
		cc3a & "	& $h_{\rm cav} = 20$ kpc, $\theta_{\rm cav} = 45^\circ$, $\theta_{\rm view} = 45^\circ$\\
		\hline
		bh1a & min(\tctf) $\approx 12$, no cavity, with black hole	& $M_{\rm bh} = 10^{10} \msun$ \\
		bh2a & "		& $M_{\rm bh} = 10^{12} \msun$ \\
		\hline
		cc1b & Conical cavity with min(\tctf) $= 5$ & $h_{\rm cav} = 20$ kpc, $\theta_{\rm cav} = 30^\circ$\\
		cc2b & "	& $h_{\rm cav} = 20$ kpc, $\theta_{\rm cav} = 30^\circ$, $\theta_{\rm view} = 45^\circ$\\
		cc3b &	"	& $h_{\rm cav} = 50$ kpc, $\theta_{\rm cav} = 45^\circ$ \\
		cc4b & "	& $h_{\rm cav} = 50$ kpc, $\theta_{\rm cav} = 45^\circ$, $\theta_{\rm view} = 45^\circ$\\
		cc5b & "	& $h_{\rm cav} = 50$ kpc, $\theta_{\rm cav} = 45^\circ$, $\theta_{\rm view} = 0^\circ$\\
		cc6b & Same as cc4b but the conical cavity is replaced by gas with $T = 7$ keV. This gas follows the HSE pressure profile & $h_{\rm cav} = 20$ kpc, $\theta_{\rm cav} =  45^\circ$, $\theta_{\rm view} = 45^\circ$\\
		\hline
		dscb & Displaced spherical cavity with min(\tctf) $= 5$ & $r_0 = 10$ kpc, $z_0 = 15$ kpc \\
		\hline
		flr  & A floor in the \tctf placed at the center & floor (\tctf) $= 10$ \\
		\hline
		pr1a & 3D geometry with perturbation, min(\tctf) $\approx 12$ & $\frac{\Delta \rho}{\rho}|_{\rm max} = 0.01$ \\
		pr2a & "	& $\frac{\Delta \rho}{\rho}|_{\rm max} = 0.1$ \\
		\hline
		simt0 & Simulation data, min(\tctf) $\approxeq 9.9$	& $t = 0$ Myr\\
		simt1000 & Simulation data, min(\tctf) $\approxeq 11$  & $t = 1000$ Myr	 \\
		simt1000th0 &  Simulation data, min(\tctf) $\approxeq 11$  & $t = 1000$ Myr,  $\theta_{\rm view} = 0^\circ$ \\
		simt170     & Simulation data,  min(\tctf) $\approxeq 2.4$     & $t = 170$ Myr \\
		simt840     &  Simulation data, min(\tctf) $\approxeq 3.08$    & $t = 840$ Myr \\
		simt1910   & Simulation data, min(\tctf) $\approxeq 3.05$    & $t = 1910$ Myr\\
		\hline \hline
	\end{tabular}
		\caption{Parameters used in the test models and simulations. The first two letters of each model name (except for `flr' and `simt' models) indicate the geometry of the model while the last letter indicates the min(\tctf) of that model (for example, `a' implies min(\tctf) $= 12$ and  `b' implies min(\tctf) $= 5$). Unless otherwise mentioned, the viewing angle, $\theta_{\rm view} = 90^\circ$ (\tt{i.e.,} edge on) by default.}
\end{table*}
\begin{figure*}
    \centering
    \includegraphics[width=0.9\textwidth, clip=true, trim={0cm, 7cm, 4cm, 9cm}]{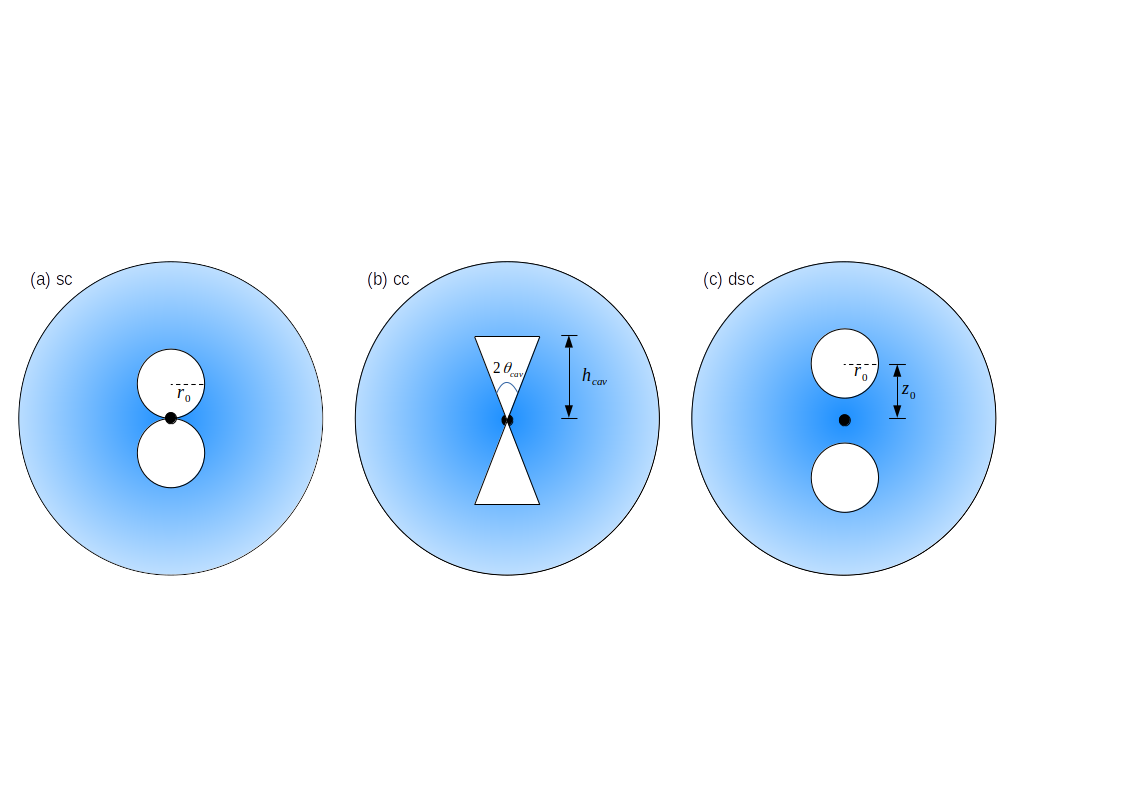}
    \caption{Cartoon diagram of different cavity models considered - (a) spherical cavity models, (b) conical cavity model, and (c) displaced spherical cavity model. The symbols contain the same meaning as in table \ref{table:test-models}.}
    \label{fig:cavity-cartoon}
\end{figure*}

\subsubsection{Spherically symmetric toy models}
\label{subsubsec:model-1d}
These are spherically symmetric models where the density and temperature distributions were obtained by considering the gas to be in hydrostatic equilibrium against the gravitational potential of a central BCG and NFW dark matter distribution. The BCG potential is taken as a singular isothermal sphere with $\Phi = V_c^2 \ln (r/a_0)$, where $V_c=350$ km s$^{-1}$ and $a_0=3$ kpc. The dark matter halo mass  $M_{200} = 7 \times 10^{14} M_\odot$ and the concentration parameter is $4.7$. The entropy profile is taken to be 
\begin{equation}
\label{eq:Kent}
K \equiv \frac{T_{\rm keV}}{n_e^{2/3}} = K_0 + K_{100} \left( \frac{r}{100~{\rm kpc}} \right)^{1.4}
\end{equation}
(motivated by \citealt{cav09}) and $n_e=7 \times 10^{-4}$ at the outer radius of 500 kpc. For more details of the setup please refer to the initial conditions and parameters described in \citet{prasad15, prasad18}.

We considered two type of models where the min(\tctf) $\approx 12$ and min(\tctf) $\approx 5$. The minimum values were obtained by adjusting the parameter $K_0$ (the core entropy; 10, 2 keV cm$^{2}$ for the two cases, respectively) in the setup described above. We use these models to test our projection-deprojection tool as well as use them as the default ICM models on which different non-hydrostatic components are added.

\subsubsection{Cavity models}
\label{subsubsec:model-cavity}
Clusters are generally believed to host X-ray cavities, thanks to the activity of an AGN at the BCG. Depending on the accretion rate, orientation and precession of the AGN, the cavities can be of different shapes and sizes. The cavities are believed to be hot but of low density and therefore they appear to be devoid of any X-ray emission. To take into account these effects, we consider a range of simplistic cavity models where the X-ray emission was manually set to be zero inside the spherically symmetric hydrostatic gas distribution. 
\begin{enumerate}
	\item \textit{Spherical}: These cavities consist of two spherical empty regions, each of radius $r_0$ and put at $z = \pm r_0$ to represent hot bubbles rising from the central BCG but has not yet disconnected from the center.
	\item \textit{Displaced spherical:} It is also possible that the AGN has shut off in the recent past leaving behind cavities that have risen up due to buoyancy and lost contact from the center. To mimic such situations, we consider two spherical cavities of radius $r_0$ centered at $(0, 0, \pm z_0)$.
	\item \textit{Conical}: These are two conical (right circular) shaped cavities, the symmetry axis of which is lying along the Z-axis with the apex being at ($0,0,0$). These cavities are characterized by their height $h_{\rm cav}$ and half opening angle $\theta_{\rm cav}$.
\end{enumerate} 

Detailed values of $r_0, z_0, h_{\rm cav}$, and $\theta_{\rm cav}$ in different models can be found in table \ref{table:test-models}. An illustration of the cavity models is also shown in figure \ref{fig:cavity-cartoon}.

\subsubsection{Black hole potential}
\label{subsubsec:model-black-hole}
Although H17 considered both BCG and NFW potential to construct the hydrostatic equilibrium of clusters, it is also  possible that 
there is an extra potential component that is not closely approximated by the NFW or the isothermal form, e.g., the potential due to the central supermassive black hole. To test this scenario, we consider two cases where a black hole of mass $10^{10} \msun$ and $10^{12} \msun$ was put at the center of the potential and no cavity was considered. While the latter mass is not a realistic case, it nevertheless can test the dependence on the assumed potential form.  
The black hole gravity is given as $g_{\rm bh} = - G M_{\rm bh}/r^2$, where, $M_{\rm bh}$ represent the mass of the black hole.

\subsubsection{With perturbation}
\label{subsubsec:model-w-pert}
 In the case of X-ray analysis, it is known that the observed spectrum biases the phase with higher emission measure (EM) in that particular spectral band.  In a realistic cluster where the ICM contains density and temperature fluctuations, the observation may get biased towards a lower temperature (implying higher density at pressure equilibrium) phase. Such situations may underestimate the cooling time (due to higher density obtained) and, therefore, underestimate the \mtctf 
 compared to the averaged values at that radius. To check the bias, we seed the ICM with perturbations with the maximum perturbation amplitude $(\Delta \rho/\rho)_{\rm max}=0.1$ and wavenumbers lying between $8 \pi/(1 \rm Mpc)$ and $40 \pi/(1 \rm Mpc)$; the form of these perturbations is somewhat arbitrary but broadly consistent with the volume-filling ICM in simulations/observations which show $\delta \rho/\rho < 1$; e.g., see \citealt{Zhuravleva2014}. The way to generate the perturbation field is described in section 4.1.2 of \citet{choudhury2016}.
 The mean value of the \tctf ratio is, however, kept to be the same as the toy models considered previously.
 
 \subsubsection{Floor in min(\tctf)}
 \label{subsubsec:floor-tctf}
 As shown by \cite{panagoulia14} and H17 that the cool-core clusters often show an entropy profile $\propto r^{2/3}$ in the central region indicating an expected floor in the \tctf profile  rather than a typical minimum \footnote{For entropy, $S\propto r^{2/3}$, the temperature, $T\propto n_e^{2/3} r^{2/3}$. Now since, $t_{\rm cool} \propto T/(n_e\Lambda) $ and $t_{\rm ff} \propto \sqrt{r/g}$, where, $g \sim P/(n_e r) \propto T/r$, it can be shown that \tctf $\propto 1/\Lambda$ which is roughly independent of the radius.} However, H17 does not find any such floor in the deprojected \tctf profiles of the clusters that show a $r^{2/3}$ entropy profile. They attributed this discrepancy to possible large density inhomogeneity towards the smaller radii. To test this hypothesis, we also consider a case where we set the central entropy profile such that it represents a floor of \tctf $= 10$. To generate this profile, we assume the same outer entropy profile as our previous models (Eq. \ref{eq:Kent} with $K_0=10$ keV cm$^2$). However, within the radius at which \tctf falls below 10, we impose \tctf$=10$ rather than the assumed entropy profile. 
 
\subsubsection{Realistic simulations}
\label{subsubsec:model-sim}
Apart from considering simplistic toy models, we also consider 3D simulation results to understand the bias in a realistic environment. A brief description of the simulation is given here, but the reader is directed to \cite{prasad15,prasad18} for more details. Basically, these idealized, isolated ICM  simulations start with initial conditions as described in \ref{subsubsec:model-1d} but evolve the ICM self consistently in presence of radiative cooling and a simple model of AGN feedback. The cluster core is regulated in form of cooling/feedback-heating cycles whose properties are governed by the feedback efficiency and the halo mass. The cores deviate from spherical symmetry during the heating phases when jets/bubbles are prominent.

Detailed values of the parameters and description of the above described models and simulations can be found in table  \ref{table:test-models}.

\begin{figure*}
	\centering
	\includegraphics[width=0.6\textwidth, angle=-90, clip=true, trim={2.5cm 1.5cm 3cm 0cm}]{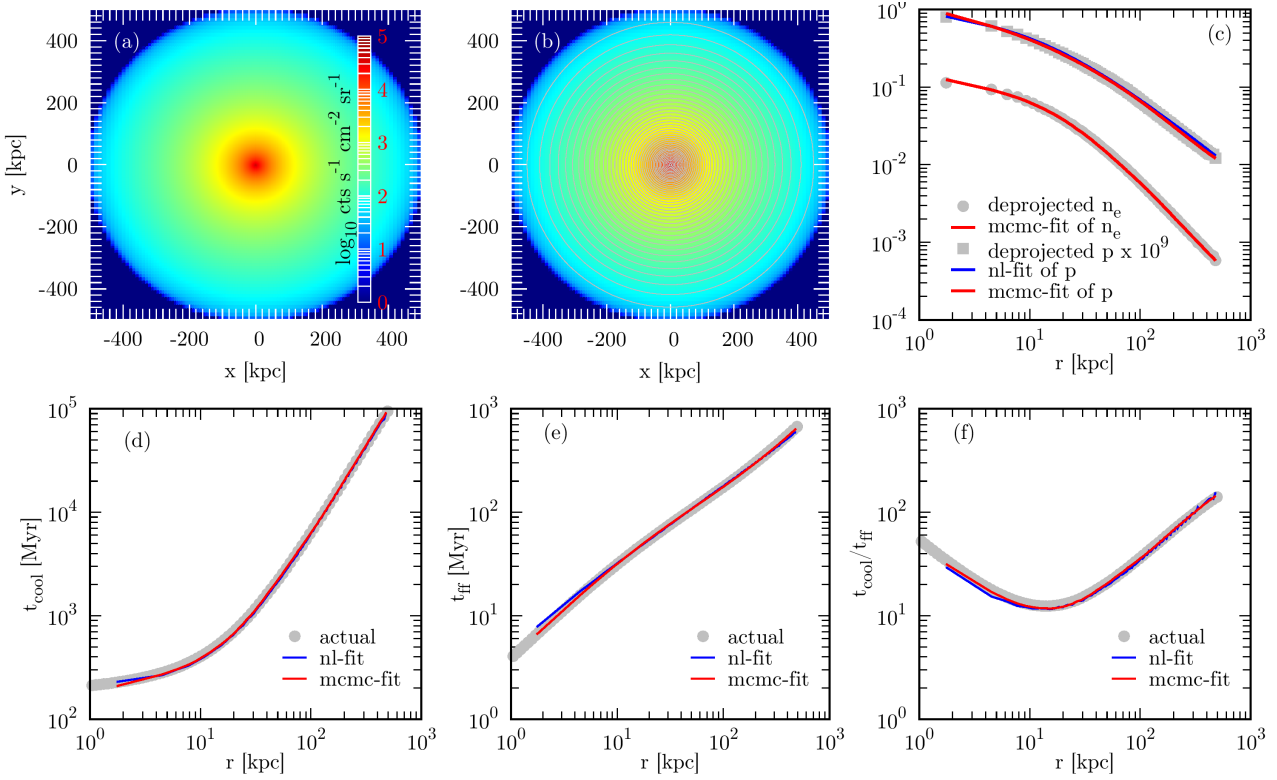}
	\caption{Testing the projection, deprojection, and reconstruction method for a 1-D (spherically symmetric) model without any cavity as described in \ref{subsubsec:model-1d}. (a) Projected surface brightness map in $2.0-14.0$ keV using \textsc{pass}. Although the brightness is given in actual units, remember that this cluster is placed at a redshift of $10^{-3}$ and therefore, appears to have an unrealistically high surface brightness. If put at higher redshifts, this cluster would have a much more realistic absolute surface brightness. (b) Annuli considered for deprojecting the spectra. We only plot every $3$rd annulus considered originally to increase visibility. (c) Deprojected electron density ($n_e$) and pressure ($p$) profiles, along with the fits obtained for them in the two fitting methods considered. The bottom panels show the inferred properties \textit{viz.}, \tc (panel d), \tf (panel e) and \tctf (panel f), and their theoretical values (gray points). } 
	\label{fig:methods-check}
\end{figure*}

\subsection{Projection tools}
\label{subsec:projection-tool}
The models above are projected using the Projection Analysis Software for Simulation (\pass\footnote{\pass is available freely from \url{https://github.com/kcsarkar}}; \citealt{sarkar17}) to obtain the surface brightness and X-ray spectrum at different lines of sight and from different angles. We assume the gas to follow \textsc{apec} plasma model \citep{apec2001}
for the emission.  Unlike the observed clusters which lie between a redshift of $\approx 0.03-0.5$, the test clusters are put at a redshift of $0.001$ to avoid limitations due to photon statistics.
A much lower redshift gives us a larger number of photons compared to the actual observations of clusters at higher redshifts, and therefore, may suffer from photon statistics (due to low photon count). It becomes necessary for the observations to keep both the energy and the spatial resolutions coarse to maintain a high  photon count in the spatial and energy bins. Therefore, for the majority of the cases we consider here, we keep the cluster at a much low redshift to remove such limitations. In section \ref{sec:photonstats} we show the implications of putting the cluster at a typical redshift.

Once projected spectra are obtained, they are then divided into different radial bins to obtain an average spectrum for each annulus. These average spectra are then used to obtain a radial profile for the electron density and temperature. In actual observations, this division is done based on a minimum number of counts per annulus so that each spectrum contains enough photons for a reliable spectral analysis. Since our test clusters are so close and bright, our spectral fits are only very slightly dependent on the total number of photons in the spectrum. 
We still consider a minimum number count per annulus as a criterion to divide the spectra into radial bins. While making the radial bins, we consider a minimum count of $10^7$ photons in each bin 
assuming an effective area, $A_{\rm eff} = 100$ cm$^2$ and for an exposure time, $t_{\rm exp} = 100$ ksec (typical for Chandra observations). The average flux is then calculated in units of photons s$^{-1}$ cm$^{-2}$ keV$^{-1}$ after multiplying with the solid angle of the radial bin. This is the flux that we fit in \xspec. While fitting in \xspec, we further assume that the instrumental response function (RSP) is a unit matrix, i.e., photon flux directly converts into counts. We use the \textit{flx2xsp} command in \xspec to generate the \textit{.pha} and \textit{.rsp} files. For the calculations presented here, we do not consider photon statistics and only assume a fixed $1$\% error in the photon counts. We study the dependence of our results on the bin sizes and realistic photon statistics in section \ref{sec:photonstats}.

\subsection{Deprojecting the data}
Once the spectra have been radially binned, they are then deprojected using a similar method as presented in \textsc{dsdeproj} \citep{dsdeproj2, dsdeproj1}.

This deprojection method is based on the assumption of spherical symmetry. As in \citealt{kriss83}, the surface brightness image is divided into concentric circular rings with decreasing radius $r_{1}, r_{2}, ...., r_{n}$ (i.e., $r_i > r_{i+1}$). The actual three dimensional cluster is assumed to be consisted of spherical shells whose radii are equal to the concentric circular rings. The volume emissivity throughout one such shell is then assumed to be constant. Therefore, the surface brightness, $S_{i,i+1}$ (cts s$^{-1}$ arcmin$^{-2}$)  and the emission density $C_{j,j+1}$ (cts $s^{-1}$ cm$^{-3}$) inside the annulus bounded by $r_{i}$ and $r_{i+1}$ depend on each other through the following relation:
\begin{equation}
\begin{split}
\label{eq:deproj}
S_{i,i+1} = \frac{b}{A_{i,i+1}} \sum_{j=1}^{i} C_{j,j+1} [V_{j,i+1} - V_{j+1,i+1}
 - V_{j,i} + V_{j+1,i}]
\end{split}
\end{equation}
where, $A_{i, i+1} = \pi (r_i^2 - r_{i+1}^2)$ is the area of the bin on the sky bounded by $r_i$ and $r_{i+1}$, $b = (\pi D/10800)^2$, $D$ is the distance to the cluster in cm and $V_{m,n} = 4\pi/3 \times (r_m^2-r_n^2)^{3/2}$ for $m < n$ and $0$ otherwise.

We can now start from the outermost annuli and find $C_{j,i}$ values for all the shells assumed. Here, we should mention that unlike \textsc{dsdeproj} we do not need to  subtract the background error because we are deprojecting modeled/simulated data.

Once the spectra of all the shells have been deprojected, we fit it with \textsc{xspec} assuming a single temperature \textsc{apec} 
model and obtain the temperature and normalization for a known metallicity. This normalization is then used to find the density of gas inside each spherical shells using the following relation 
\begin{equation}
    \label{eq:norm_to_density}
    \text{Norm} = \frac{10^{-14} \int n_e n_H dV}{4 \pi (D_A [1+z])^2} = \frac{10^{-14} n_e n_H \int dV}{4 \pi (D_A [1+z])^2},
\end{equation}
where $z$ is the redshift, $D_A$ is the \textit{angular diameter distance}, $n_e$ and $n_H$ are electron and hydrogen densities in cm$^{-3}$, respectively. We use these quantities in the next step of our analysis.

\subsection{Reconstruction}
After the deprojected density and temperature profiles have been obtained, we can easily estimate the cooling time as
\begin{equation}
\label{eq:tcool}
t_{\rm cool} \approx \frac{3}{2}\: \frac{n k_B T}{n_e n_i \Lambda(T, Z)}\,,
\end{equation}
where $\Lambda(T, Z)$ is the cooling efficiency taken from \cite{sutherland+dopita+93} \footnote{More recent cooling curves, such as in \textsc{cloudy}-17 \citep{cloudy2017} show significantly lower cooling rates for $T \lesssim 10^6$ K and for lower metallicity (see figure \ref{appfig:cooling-comparison}). This discrepancy, however, is only $\approx 1.5$ for $10^6-5\times 10^6$ K gas, above which the discrepancy is even lower. Thus, considering new cooling rates will not change our results much since the plasma in our toy models and simulations has temperature $T\gtrsim 10^7$ K.}, $n = \rho/(\mu m_p)$ is the total particle density, $n_e = \rho/(\mu_e m_p)$ is the electron number density, $n_i = \rho/(\mu_i m_p)$ is the total ion number density; we choose $\mu = 0.598$, $\mu_e = 1.151$, $\mu_i = 1.248$. We take these parameters to be constant throughout our calculation. In reality, the values of $\mu$, $\mu_e$, and $\mu_i$ remain practically constant for $T \gtrsim 10^5$ K gas and for a large range of metallicities. The metallicity is considered to be $0.2$ Z$_\odot$ for all the toy models including the simulations. We also show in figure \ref{fig:error-NL} that we are able to recover the profiles quite well even for a higher metallicity. A higher metallicity generally decreases the cooling time of the ICM but this decrease is not more than by a factor of $2$ for $T\gtrsim 1$ keV (see fig \ref{appfig:cooling-comparison}).

The free-fall time is given by
\begin{equation}
\label{eq:tff}
t_{\rm ff} \approx \sqrt{\frac{2\:r}{g}} \,,
\end{equation}
where $g$ is the gravitational acceleration at radius $r$. However, obtaining \tf requires the assumption of hydrostatic equilibrium in the cluster despite the presence of cavities and non-axisymmetric features in many of the observed clusters. 

We follow the traditional methods to obtain \tf from the deprojected profiles to test the accuracy of these methods. We consider two methods to obtain $g$ for each of the scenarios. In the first method, the deprojected pressure ($p = n\:k_B\: T$; $n = \mu_e n_e/\mu$) is fitted with a broken power law of the form
\begin{equation}
P(r) = \frac{P_0}{\left( \frac{r}{a}\right)^{\alpha_1} + \left( \frac{r}{a}\right)^{\alpha_2}}
\label{eq:nl-pressure}
\end{equation}
and then the derivative of the pressure is used to obtain the gravitational acceleration since in hydrostatic equilibrium,
\begin{equation}
g(r) = - \frac{1}{\rho}\: \frac{d\:P}{dr}\,.
\label{eq:nl-gr}
\end{equation}
\citep{lakhchaura16}. This method will be referred to as \textit{nl-fit} (nonlinear fit) from here on.

The second method follows the reconstruction of the hydrostatic equilibrium in each of the deprojected shells by assuming a global $g(r)$ profile for the dark matter and a known BCG. Considering the temperature of the shell and assuming pressure continuity across the shell boundaries, the hydrostatic equilibrium can be solved given the density at the inner shell boundary. 
For example, if the temperature of a shell bound by $r_i$ and $r_{i+1}$ is taken to be a constant ($T_i$) and the density at $r_i$ is $n_i$, then given a gravitation acceleration $g(r)$, one can write down the density at $r_{i+1}$ to be
\begin{equation}
 n_{i+1} = n_i \exp\left( -\frac{\mu m_p}{k_B T_i} \int_{r_i}^{r_{i+1}} g(r) dr \right) \,.
\end{equation}
Note that, although the pressure remains continuous across the shell boundaries, the density and temperature (which is stepwise constant) do not. Once the pressure and the density profiles have been reconstructed, the free parameters of the gravitational potential, \textit{viz.} the virial mass $M_{200}$ and the concentration parameter $c$ can be found by fitting the obtained density profiles to the deprojected ones. The gravitational acceleration, $g(r)$, is assumed to be a combination of the dark matter profile (assumed NFW), a central BCG profile and acceleration due to the SMBH (if any). The NFW gravity profile is taken to be
\begin{equation}
g_{\rm nfw} (r) =  - \frac{G\:M_{200}}{f(c)\: r_{200}^2} \times \frac{1}{x^2}\:\left[ \log(1+c\:x) - \frac{c\:x}{1+c\:x} \right]
\label{eq:g-nfw}
\end{equation}
where, $x = r/r_{200}$, $r_{200} = \left( \frac{3 M_{200}}{200\times 4\pi \rho_c}\right)^{1/3}$ is the virial radius and $\rho_c = 8.64 \times 10^{-30}$ g is the critical density of the universe at redshift zero, and $f(c) = \log (1+c) - c/(1+c)$. From here onward, we will call this method as \textit{mcmc-fit} (since the parameter space of gravitational potential is explored using MCMC). This method is similar to the one followed by \cite{hogan17} except that we use the BCG gravity to be 
\begin{equation}
\label{eq:vc}
g_{\rm bcg} = -\frac{v_c^2}{r_\ast}
\end{equation}  
where, $r_\ast = \sqrt{r^2+(0.2\:{\rm kpc})^2}$ and $v_c = 350$ \kmps by construction.
We also use, in some cases, an additional acceleration from the central SMBH in the form of 
\begin{equation}
    g_{\rm bh} = - \frac{G\:M_{\rm bh}}{r_\ast^2} \,,
\end{equation}
where $M_{\rm bh}$ is the mass of the SMBH.

\subsection{Testing the method}
\label{subsec:method-testing}
Figure \ref{fig:methods-check} shows a step-by-step procedure (from $a$ to $f$) of our analysis performed in this paper. It shows results for a spherically symmetric atmosphere in HSE without any cavity/perturbations (as described in section \ref{subsubsec:model-1d}). We find that both the methods (\nlfit and \mcfit) are equally good at fitting the observed pressure profiles and obtaining the cooling time \tc. \mcfit is, however, better at reproducing the \tf profile towards the center. This is expected since the total potential at very small radii is dominated by the BCG potential which is a  
parameter for \mcfit but is unknown to \nlfit. In observations, the BCG potential is obtained by estimating the stellar mass of the galaxy \citep{hogan17, pulido18}. In any case, we note that both the \nlfit and \mcfit are successfully  able to retrieve the original profiles to very good accuracy.  We, therefore, proceed with the analysis of the toy models/ simulation data as listed in Table \ref{table:test-models}.

\begin{figure*}
	\centering
	\includegraphics[width=0.9\textwidth, clip=true, trim={0cm 3cm 0cm 2cm}]{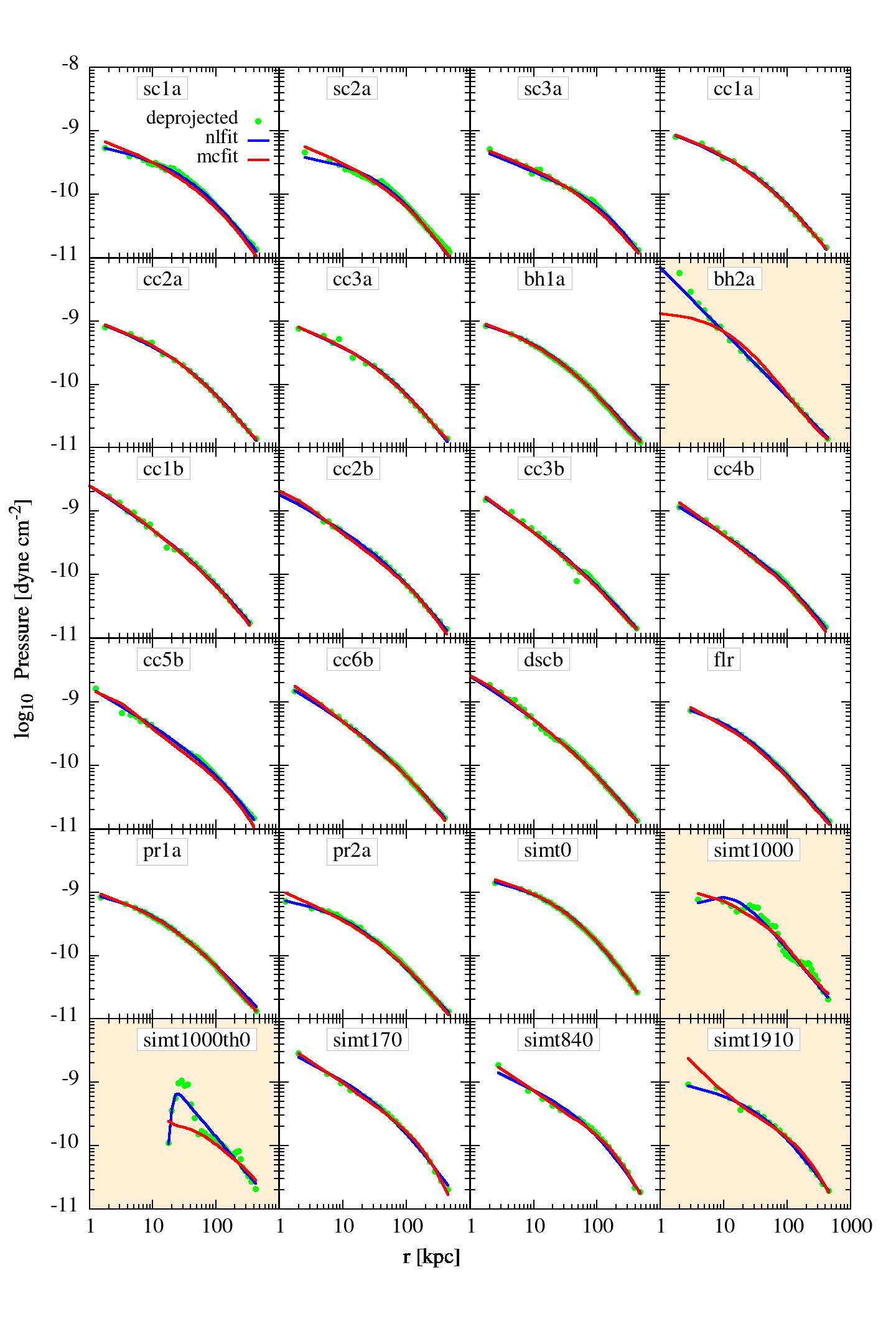}
	\caption{Best fit to the deprojected pressure data points (green points). The  solid blue and red lines represent \nlfit and \mcfit, respectively. Cases that are not fitted well are shown with an orange background. We keep the same orange background in the next few figures for the results obtained from these cases. Note that the effect of cavities in the deprojected `sc' models appears at $r < 2 r_0$ because the cavities are placed at $z = \pm r_0$. }
	\label{fig:pressure-fit}
\end{figure*}

\begin{figure*}
	\centering
	\includegraphics[width=0.9\textwidth, clip=true, trim={0cm 3cm 0cm 2cm}]{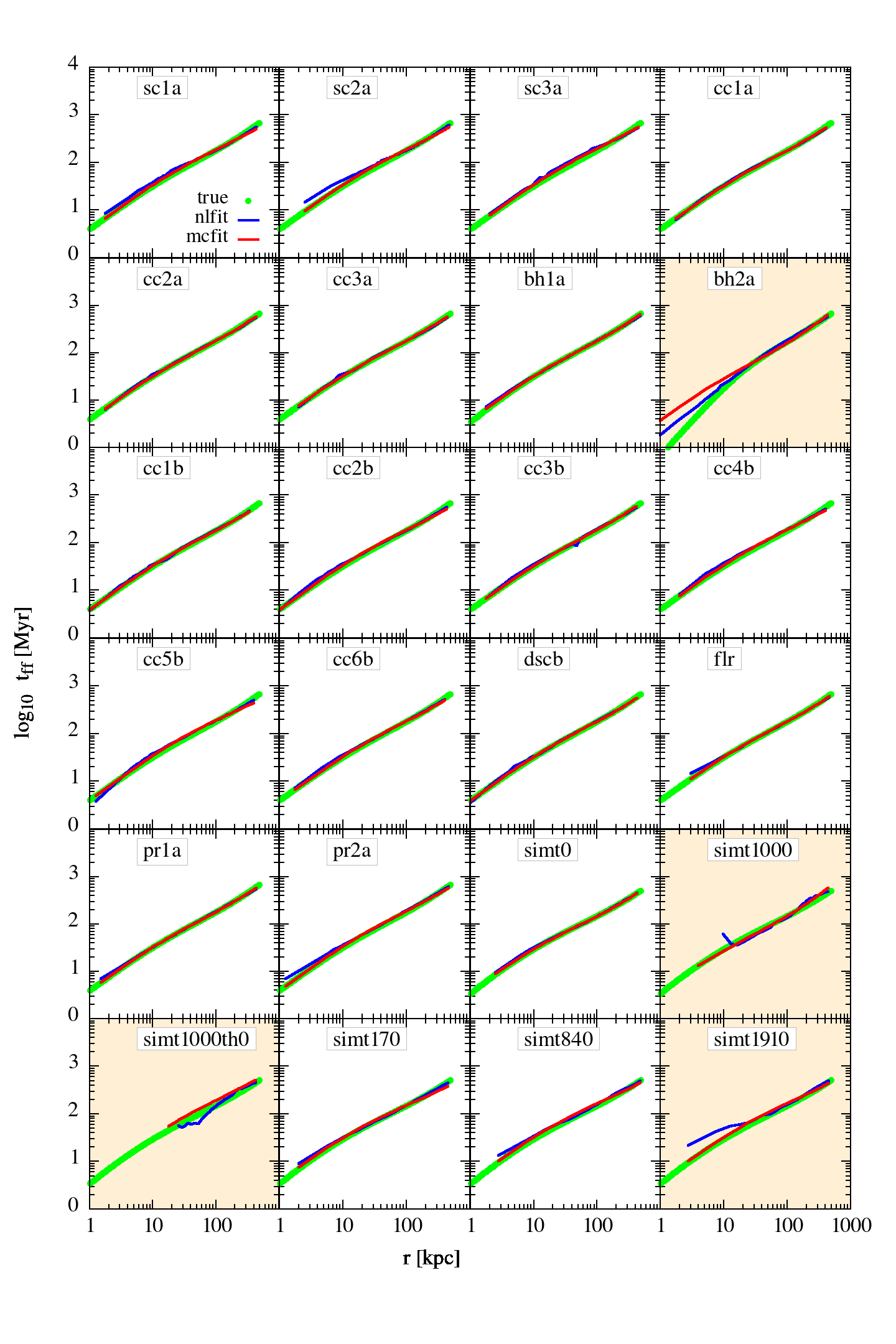}
	\caption{Reconstruction of $t_{\rm ff}$ (equation \ref{eq:tff}) using the best fit parameters from both the fits. The red and blue lines show the reconstructions for \tt{mlfit} and \tt{nlfit}, respectively. 
    The green circles represent the true free-fall time estimated from the known mass of the toy models/simulations. The panels with orange background represent slightly poor fits as shown in figure \ref{fig:pressure-fit}.}
	\label{fig:tff}
\end{figure*}
\begin{figure*}
	\centering
	\includegraphics[width=0.9\textwidth, clip=true, trim={0cm 3cm 0cm 2cm}]{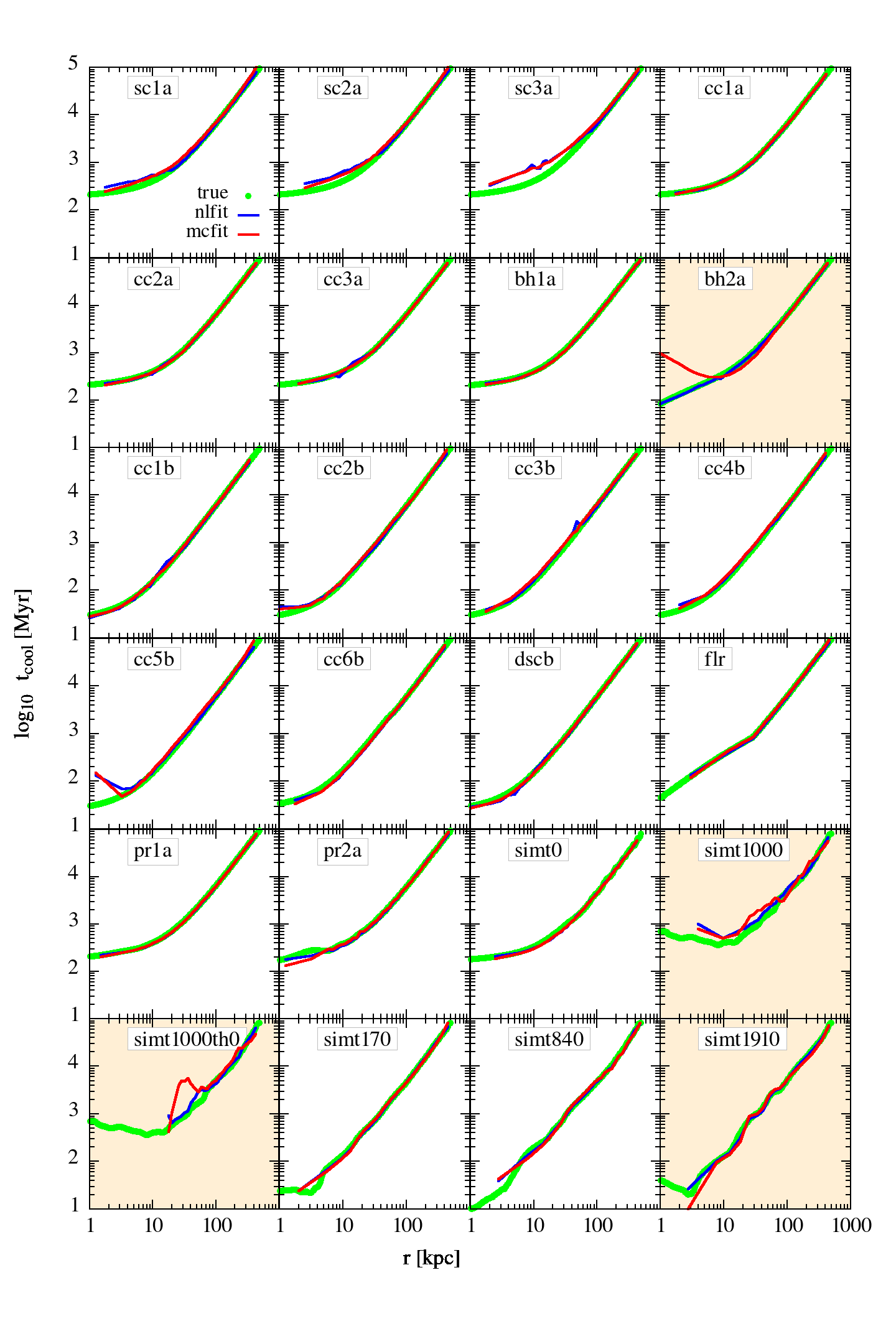}
	\caption{Reconstruction of \tc (equation \ref{eq:tcool}) using the deprojected density and temperatures (in case of \nlfit) or using the best fit parameters (in case of \mcfit). The blue and red lines represent results from \tt{nlfit} and \tt{mcfit} methods, respectively. 
	The green circles are the estimated cooling time from the toy models/simulations as described in Sec. \ref{subsec:recovering-tc}. The panels with orange background represent cases that have poor fits as shown in fig \ref{fig:pressure-fit}.}
	\label{fig:tcool}
\end{figure*}

\begin{figure*}
	\centering
	\includegraphics[width=0.9\textwidth, clip=true, trim={0cm 3cm 0cm 2cm}]{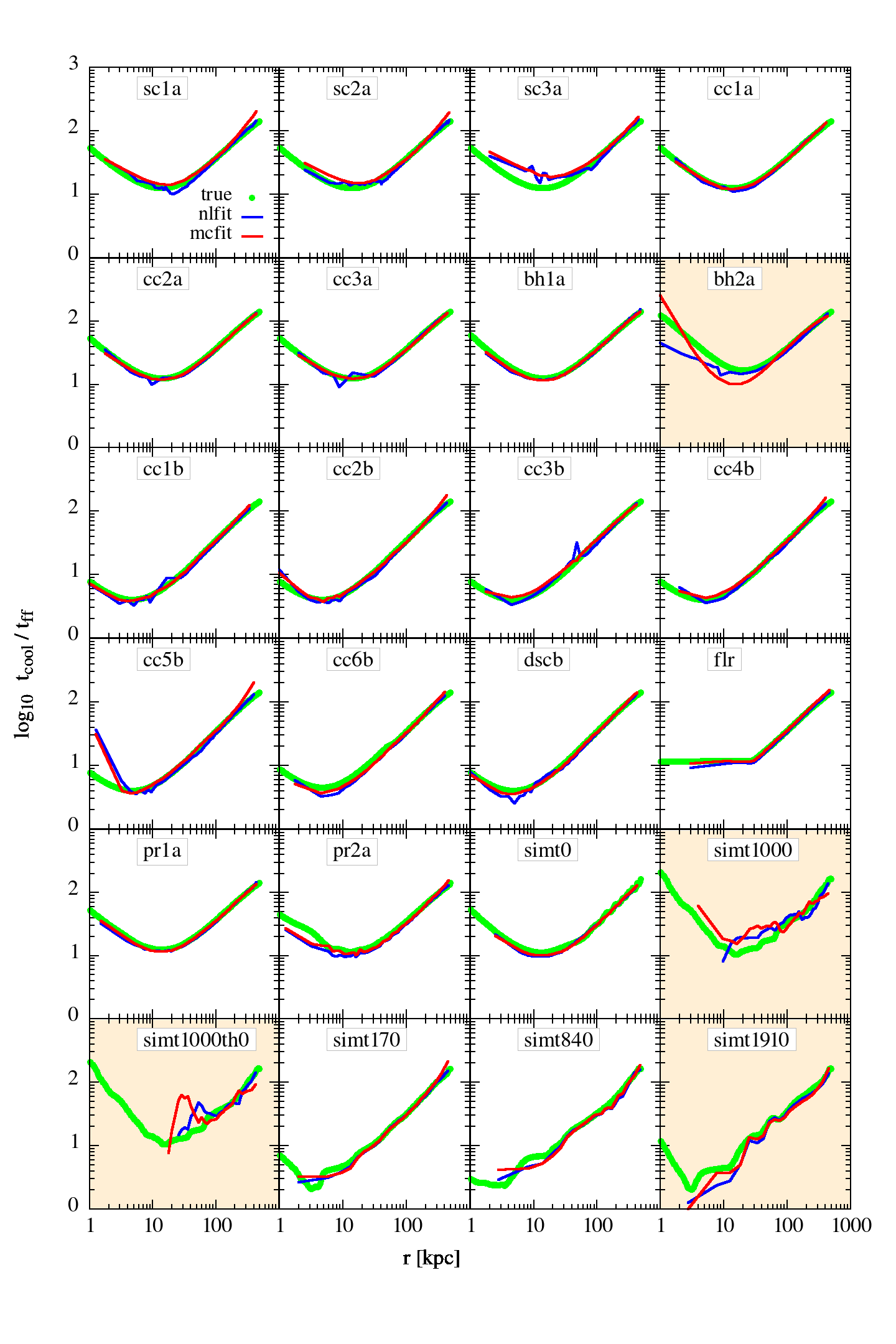}
	\caption{Reconstruction of \tctf using the best fit parameters from \tt{nlfit} (blue line) and \tt{mcfit} (red line). 
	The green points are the true \tctf ratio of the toy models/simulations.}
	\label{fig:tcool-tff}
\end{figure*}

\section{Results}
\label{sec:results}

\subsection{The fits}
\label{subsec:results-fits}
We present results of the fits in Fig. \ref{fig:pressure-fit} for all the cases mentioned in table \ref{table:test-models}. Note that the pressure is not directly fit in \mcfit; it only fits the density and the pressure is obtained by multiplying the temperature and density profiles. 
We notice that the fits are good in both the methods in most of the cases except for tiny differences at very  
small radii. This difference is due to the presence of cavities 
in the core that cause the pressure 
there to become non-smooth and even hard to fit in some cases. The difference, however, is prominent in \textit{bh2a, simt1000, simt1000th0} and \textit{simt1910} cases (marked by orange background in the figure). In the case of \tt{bh2a} (with a $10^{12} M_\odot$ point mass at the center), the \nlfit works better as it does not assume a parametric form for the fitted potential and can easily fit an unknown potential.
This would also be true for a BCG in case its mass is known inaccurately. 
In \tt{simt1000}, although the fits are not very good, they roughly recover the trend. The one case, where \mcfit fails completely is \tt{simt1000th0} (viewing angle along the jet axis) where the deprojected density and hence the pressure goes to zero for $r \lesssim 20$ kpc. This is because, for such a projection, the bubble appears to be a hole at the center which, therefore, causes the deprojection method (Eq. \ref{eq:deproj}; which assumes spherical symmetry) to fail as the surface brightness decreases towards the center instead of increasing. 
Since the inner slope in \mcfit is fixed  
to a known BCG potential \tt{i.e.} a known $v_c$ (equation \ref{eq:vc}), it does not follow the negative slope of the obtained pressure. On the other hand, the \nlfit method being independent of the form of the potential recovers the pressure trend. As we will see later, this badly affects the ability for \mcfit to estimate \tc. A similar reasoning can also be attributed to the bad fitting of \mcfit method in \tt{bh2a} case. However, the value of the black hole mass considered in this case is very large compared to realistic SMBHs which usually have masses $\lesssim 10^{10} \msun$. This example illustrates the fallibility of the \mcfit methods when the actual potential differs qualitatively from the assumed form.

\subsection{Recovering \tf}
\label{subsec:results-tff}
We also are able to recover the free-fall time, \tf, for most of the cases (see Fig. \ref{fig:tff}) quite well. As mentioned earlier, the \mcfit method retrieves the \tf better since we use 
the correct potential for the BCG in this method.
The \nlfit method is also able to recover the free-fall time to a good accuracy except in the cases where the deprojected pressure does not follow the  
assumed pressure profile very well, \tt{i.e.}, when the pressure distribution is disturbed by turbulence or when
there is a big cavity in the central region. This is because a central cavity makes the deprojected pressure profile flatter than  
the correct profile. This leads to an underestimation of gravitational acceleration ($g$) and, therefore, an overestimation of \tf since \tf $ = \left( - \frac{2\:\rho \: r}{dP/dr}\right)^{1/2}$, for example, in \tt{sc2a} and \tt{sim1910}.  Although, \tt{sc3a} has a cavity size ($40$ kpc) bigger than \tt{sc2a} ($20$ kpc), it recovers \tf better.
This is mainly due to a poor pressure fit in \tt{sc2a} which shows a more prominent kink in the deprojected pressure, leading to a shallower pressure fit.

The figure also shows that, in case of an unexpected gravitational potential at the center, such as
\tt{bh2a}, profiles are better fitted by the \nlfit method. The \mcfit method fails to account for the extra potential in such cases. A way to get around this issue in this method could be to allow fitting for $v_c$. The \nlfit, however, proves to be a less effective method at recovering the \tf in the simulated cavity situations (\tt{sim1000, simt1000th0, sim1910}) than \mcfit method. We conclude that while estimating \tf, we can use \mcfit in cases where the central potential is known but should 
switch to \nlfit in the cases where we have limited information on the central potential.

\subsection{Recovering \tc}
\label{subsec:recovering-tc}

We use the definition of the cooling time as given in equation \ref{eq:tcool}.
Note that, while \nlfit considers the deprojected $n_e$ and $T$ for calculating the \tc, \mcfit considers the best fit $n_e$ and the deprojected temperature for this purpose. As we will see later that this is the reason that \nlfit can provide a better estimate of \tc than \mcfit. 

We also estimate the average cooling times from the toy models/simulations 
to compare with the values retrieved from the fitting methods. We estimate the emission weighted density and temperature at each radius and then use these values to obtain the `actual' \tc from Eq. \ref{eq:tcool}. The average density and temperatures are calculated as 
\begin{eqnarray}
\label{eq:n-avg}
n_{\rm avg} (r) &=& \frac{\int_{\theta, \phi} n \times n_e n_i \Lambda(T,0.2)\: dV}{\int_{\theta,\phi} n_e n_i \Lambda(T,0.2)\: dV}  \nonumber \\
T_{\rm avg} (r) &=& \frac{\int_{\theta, \phi} T \times n_e n_i \Lambda(T,0.2)\: dV}{\int_{\theta,\phi} n_e n_i \Lambda(T,0.2)\: dV} 
\end{eqnarray}
It is clear from the definitions of the $n_{\rm avg}$ and $T_{\rm avg}$ that the cavities containing very low density plasma do not contribute to the averaged quantities.\footnote{While calculating the average quantities, we consider that gas with $T< 0.5$ keV has zero emissivity to avoid any contamination by the cold/warm gas in the estimated quantities. This is done keeping in mind that our method of analysis only uses X-ray data.} The definitions are also consistent with the definition of \tc used in the papers  
reporting numerical simulations of AGN-driven feedback in clusters \citep[for example, ][]{prasad18}.

The `actual' \tc and the retrieved profiles are shown in Fig.  \ref{fig:tcool}. The green points represent the actual \tc, whereas, the solid blue (red) line represents the profiles obtained using the \nlfit (\mcfit) method. Clearly, both methods reproduce the actual \tc profiles quite well in most  
cases. However, \nlfit retrieves \tc better than \mcfit  in general, specifically in the cases where the fit to the deprojected pressure is not good for \mcfit \tt{viz. bh2a, sim1000th0} and \tt{simt1910}. The \nlfit, however, shows small bumps in \tc due to the direct use of deprojected density instead of a fitted density.   
In the presence of a spherical cavity, \tc is overestimated by both methods - the bigger the cavity, the larger is the discrepancy. This is  
because in observations the density is obtained from the total surface brightness of an annulus and assuming that the emissivity throughout the spherical shell is constant (see \ref{eq:deproj}).  This leads to an overall lower density of the shell to match the lower surface brightness. On the other hand, the `actual' \tc obtained from the models/data using Eq. \ref{eq:n-avg} excludes any cell that has very low emissivity. Therefore, while the information of a cavity is imprinted in the density estimation of \nlfit or \mcfit, the `actual' \tc is not much affected by the presence of cavities. This is the reason that the \nlfit or \mcfit overestimates \tc. 

Surprisingly, conical cavity models and the displaced spherical cavity model do not show such an overestimation of the \tc. This is perhaps due to a small solid angle subtended by these cavities leading to a much lower volume occupied by them at any radius compared to the spherical cavities. Since the spherical cavities are connected at the center of the cluster, they can take up almost all the volume at  
small radii and hence produce relatively low surface brightness. It is, therefore, interesting to also note that for any cluster where the AGN is switched off for a considerable time so that the cavities have risen to a significant distance, the cavities do not much affect the measurement of the \tc. Now, since \tc$\propto 1/n$ and that the average value of the density due to the cavity is given as $n_{\rm avg}/n = \cos\theta_{\rm cav}$ (in an approximate volume averaged sense), one can roughly estimate the effect of such cavities in real observations as $t_{\rm cool, obs}/t_{\rm cool} =  n/n_{\rm avg} = 1/\cos\theta_{\rm cav}$ at any given radius. For example, in case of \tt{sc3a} at $r = 10$ kpc, we estimate that $\theta_{\rm cav} \approx 70^\circ$ and, therefore, $t_{\rm cool, obs}/t_{\rm cool} \approx 3$, roughly consistent with the recovered \tc (see fig \ref{fig:tcool}). On the other hand, the same argument leads to $t_{\rm cool, obs}/t_{\rm cool} \approx 1.3$ in the case of \tt{cc3b} where $h_{\rm cav} = 50$ kpc and $\theta_{\rm cav} = 45^\circ$. 

\subsection{Recovering \tctf}
\label{results-tctf}
In fig. \ref{fig:tcool-tff}, we plot the recovered \tctf from both the methods and compare them with the `actual' \tctf profiles obtained from the toy models/simulations. We discuss the results for each subcategory  in the following sub-subsections.

\subsubsection{Cavity models}
\label{subsubsec:cavity-models}
 As pointed out earlier that both the analysis methods (\nlfit and \mcfit) are able to recover  
 the free-fall time (\tf) in most cases but overestimate the cooling time (\tc) in some cases where there is a cavity 
 subtending a large solid angle present inside the cluster. Clearly, the \tctf ratio is also  
 overestimated in those cases. The most prominent example of such a case is the \tt{sc3a} (spherical cavity of size $40$ kpc) where min(\tctf) is overestimated  almost by a factor of $2-3$ . In another example, the methods also overestimate the \tctf values within $\lesssim 5$ kpc in \tt{cc5b} (conical cavity of size $50$ kpc and viewed down the barrel \tt{i.e.}, $\theta_{\rm view} = 0^\circ$) due to the projection effects that create a hole in the surface brightness profile. 
 This causes the deprojection method to perform poorly in recovering the density and temperature profiles towards the center. Fortunately, the effect is only noticeable in the very first bin, and therefore, does not affect the minimum value of the \tctf which occurs at $r \sim 10$ kpc. Based on the discussion in sections \ref{subsec:recovering-tc} and \ref{subsec:results-tff}, we conclude that, in the presence of cavities, the error in  \tctf ratio is mostly driven by the error in \tc. The recovered \tctf ratio differs from the actual \tctf if the solid angle subtended by the cavity at the center is large.

\subsubsection{Extra potential}
As noted in the previous paragraph that the obtained \tctf profiles are mostly affected by the estimates of \tc than \tf, since \tf is well estimated in most of the cases.  One particular toy model, where the estimation of both the \tc and \tf fails in \mcfit method is \tt{bh2a}. Therefore, the \tctf profile also 
deviates from the true one in this case. On the other hand, \nlfit performs much better  
for this case and recovers the min(\tctf) value but deviates from the actual profile within $\lesssim 10$ kpc mostly due to the pressure profile not being a simple double power-law. We should, however, keep in mind that such a  
massive black hole is not realistic. Any extra potential, if it exists inside the BCG, will either be small (like \tt{bh1a}) where the \tctf profiles are recovered very accurately, or will manifest its presence in the observed $v_c$ of the BCG (a parameter in the BCG potential). However, in case the extra potential has a separate  profile than the assumed profile of the BCG (either in addition to or instead of), \nlfit provides a better way to estimate \tctf.

\begin{figure*}
    \centering
    \includegraphics[width=0.95\textwidth,clip=true, trim={2cm 0cm 2cm 1cm}]{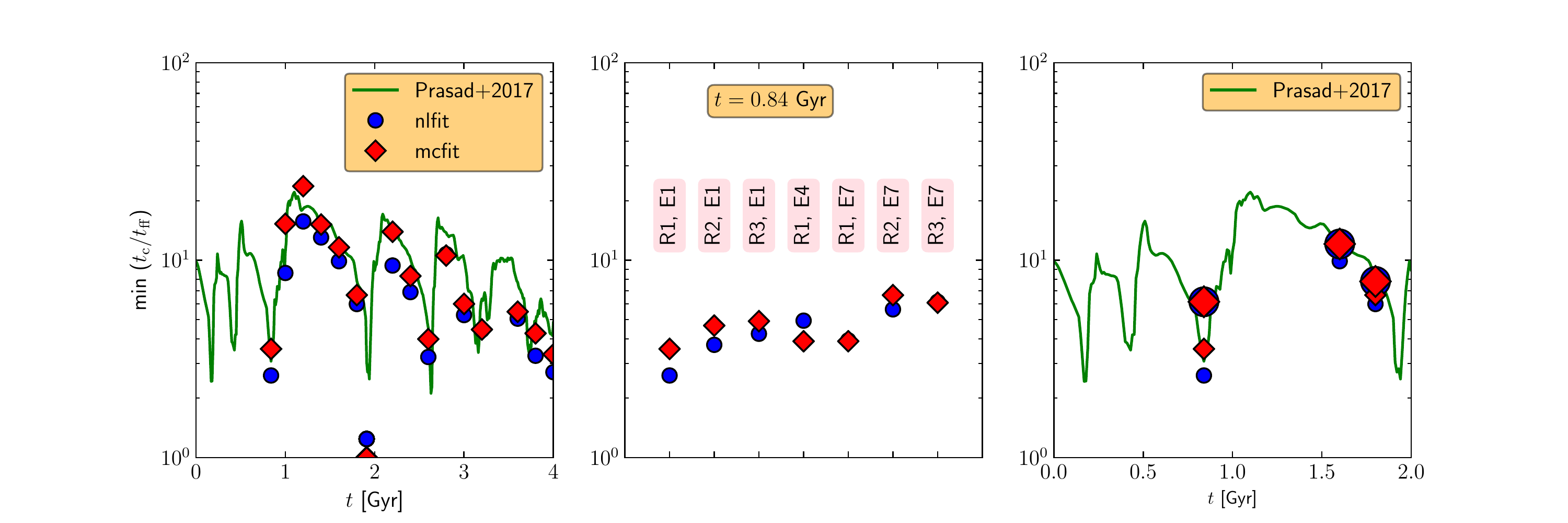}
    \caption{Comparing the min(\tctf) recovered from the \textit{nlfit} and \textit{mcfit} fits with the angular averaged values from the simulation. \textit{Left panel}: min(\tctf) values obtained at different snapshots of the simulation from theory (green solid line), 
    \textit{nlfit} (blue circles), and \textit{mcfit} (red diamonds). \textit{Middle panel}: min(\tctf) values at $t = 0.84$ Gyr for different cases with changing the observed bin-size or the observational spectral energy band. The labels R1 (default resolution), R2, R3 represent the central bin sizes of $2.5, 5.5$, and $10$ kpc, respectively, and the energy bands E1, E4 and E7 represent $0.1-15$ keV, $0.4-10$ keV, and $0.7-12$ keV, respectively. \textit{Right panel}: the effect of raising the lower energy bound of the observed spectra (shown by larger points). The very low min(\tctf) values seem to rise by a factor of $2-3$, whereas, higher min(\tctf) values do not increase significantly.}
    \label{fig:tctf_comp}
\end{figure*}

\subsubsection{With density perturbations}
\label{subsubsec:results-pert}
The toy models we considered earlier, have a well defined minimum in the \tctf value. Real cool-core systems, on the other hand, may contain perturbations over a smooth distribution either due to the turbulence induced by AGN jets/bubbles or simply due to the non-relaxed nature of the cluster. As a result, these systems may have large density/temperature variations. We mimic such systems with the \tt{pr1a} and \tt{pr2a} models, the results of which are
shown in Fig. \ref{fig:tcool-tff}, marked as \tt{pr1a} and \tt{pr2a}. Since the perturbations are added as fluctuations in density (and, therefore, in temperature), which affect the emission measure most, it is not certain that the \tctf profile will remain the same. For a better understanding of the systems, we can look into the expected change in  
\tc (and hence to  
\tctf) due to such perturbations. Since the perturbations are added at each radius for a given constant pressure, $n T = n_{\rm avg} T_{\rm avg}$ at every radius, $n = n_{\rm avg} (1\pm \delta)$, and $T$ are just the local density and temperature. Here, $\delta$ is the local density fluctuation. The cooling time therefore, now changes to \tc $\propto T^{1/2}/n$ (assuming free-free cooling). This allows us to write down the local cooling time as
\begin{equation}
 \frac{t_{\rm c}}{t_{\rm c, avg}} = \frac{n_{\rm avg}}{n} \times \left(\frac{T}{T_{\rm avg}}\right)^{1/2} \approx 1\pm 3 \frac{\delta}{2} \,.
\end{equation}
For example, since the maximum fluctuation of density in \tt{pr2a} is $0.1$, we expect about 15\% fluctuation on \tc too. Therefore, the maximum deviation of the retrieved \tc (from the `actual' \tc~) should also be of a similar order.

Figure \ref{fig:tcool-tff} shows that
both the reconstruction methods are able to recover the actual \tctf profiles for these models to a great extent. Although \nlfit underestimates \tctf in \tt{pr2a} (containing large fluctuations), this is entirely due to the overestimation of the \tf by this method. The \mcfit, on the other hand, performs much better to estimate the ratio due to the known potential at the center. We, therefore, conclude that as long as the perturbations in the system are small, both the methods work quite well. The recovered \tctf, however, may be affected only if the amplitude of density fluctuations is $\gtrsim 1$.

\subsubsection{Recovering a floor}
As mentioned in section \ref{subsubsec:floor-tctf}, we want to check if we could find a floor in \tctf (if it is present) in the core. We find that both the methods are  
able to recover the floor in \tctf (panel \tt{flr} in Fig. \ref{fig:tcool-tff}). Surprisingly, H17 does not find such a floor in their data despite having clusters with entropy $\propto r^{2/3}$  (an indication of a floor in \tctf) in the inner region, instead, they find a slight upturn in the \tctf profile at smaller radii (their figure 7). They argued that the absence of the \tctf floor may be because of the increasing impact of the density inhomogeneity towards the center. In the current analysis we find that the both \nlfit and \mcfit methods are not affected much by the presence of slight density fluctuations but can be seriously affected if the density inhomogeneity is $\gtrsim 1$. However, even if they are affected, the effect has been to flatten out rather than upturn the \tctf profile at small radii (see panel \tt{pr2a} in figure \ref{fig:tcool-tff} and related discussion in section \ref{subsubsec:results-pert}). Therefore, it seems very likely that both the methods would pick up the floor if it is present. This, therefore, means that there is probably no such floor present in the H17 data and that the clusters do not necessarily have to have a floor in \tctf even if they have entropy $\propto r^{2/3}$.

Although the above conclusion seems puzzling, the discrepancy can be understood in the following manner.  As described by H17 (their section 7), \tctf $\propto 1/(T^{1/2} \times \Lambda)$ for an entropy profile, $K \propto r^{2/3}$ (both \tc and \tf are proportional to radius, latter because the potential is close to an isothermal sphere). Here, $\Lambda$ is the cooling efficiency in units of \ergps cm$^3$. We see that 
in this case \tctf does not directly depend on the radius. 
However, if the temperature has a radial dependence, it also gets imprinted on the \tctf profile. Therefore, we speculate that the absence of a floor in the observed \tctf ratio despite the presence of $K \propto r^{2/3}$ entropy profile indicates a temperature variation in that region. It is well known that the temperature decreases towards the center in cool cores. So if the cooling function is flat or has a positive slope (as expected for free-free emission), the \tctf ratio is expected to increase inwards in the very center of cool cores.

\subsubsection{Simulated cases}
In the simulated cases, as already mentioned, although the \mcfit recovers the \tf better than \nlfit, it suffers from poor performance in recovering the cooling time-scale.  The overall \tctf is still overestimated for some cases in both the methods. As shown in case of \tt{sc3a}, we also find a similar effect of cavity in \tt{sim1000} and even more in \tt{simt1000th0}. Surprisingly, in the cases where the `actual' \tctf profile is expected to fall below $5$ (cases, \tt{simt170, simt840} and \tt{simt1910}), both the methods are able to recover the profile, although not all the way till the center (due to the  resolution limit).
This may be due to the absence of large cavities since the AGN has not started yet and the cluster is still cooling. We speculate that the effect of cooling phase at the center would be to increase the density and temperature fluctuations, similar to case \tt{pr2a} and, therefore, to affect \tctf in a similar way.

A better comparison of both the methods (\nlfit and \mcfit) can be seen in the left panel of figure \ref{fig:tctf_comp} which compiles the min(\tctf) obtained from both these methods and the actual values from simulations. The figure re-iterates the earlier point that both the methods do recover the min(\tctf) quite well throughout the duty cycle of an AGN.
This means that the discrepancy between the observed and simulated clusters, namely that the simulations show more instances of \tctf $\lesssim 10$, may not be due to any bias present in the analysis techniques. However, there may be several other effects, such as the resolution of the central bins and the inability to observe cooler gas, that can affect the estimation of the min(\tctf). We address these issues in section \ref{sec:discussion}.

\begin{figure}
	\centering
	\includegraphics[width=0.47\textwidth, clip=true, trim={2cm 1cm 5cm 3cm}]{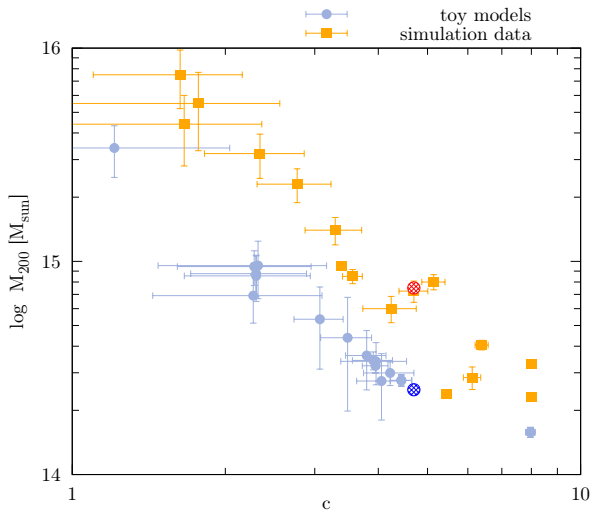}
	\caption{Retrieved virial mass $M_{200}$ and concentration parameter ($c$) from \mcfit. The blue hashed circle shows the original value used for these toy models and the red circle represents the value used in the simulation setup. A degeneracy in $M_{200}-c$ is clearly visible in both toy models and the simulation. This is discussed more in section \ref{subsec:recovering-mass}.}
	\label{fig:M200-c}
\end{figure}
\begin{figure}
	\centering
	\includegraphics[width=0.5\textwidth, clip=true, trim={0cm 0.5cm 0.5cm 0cm}]{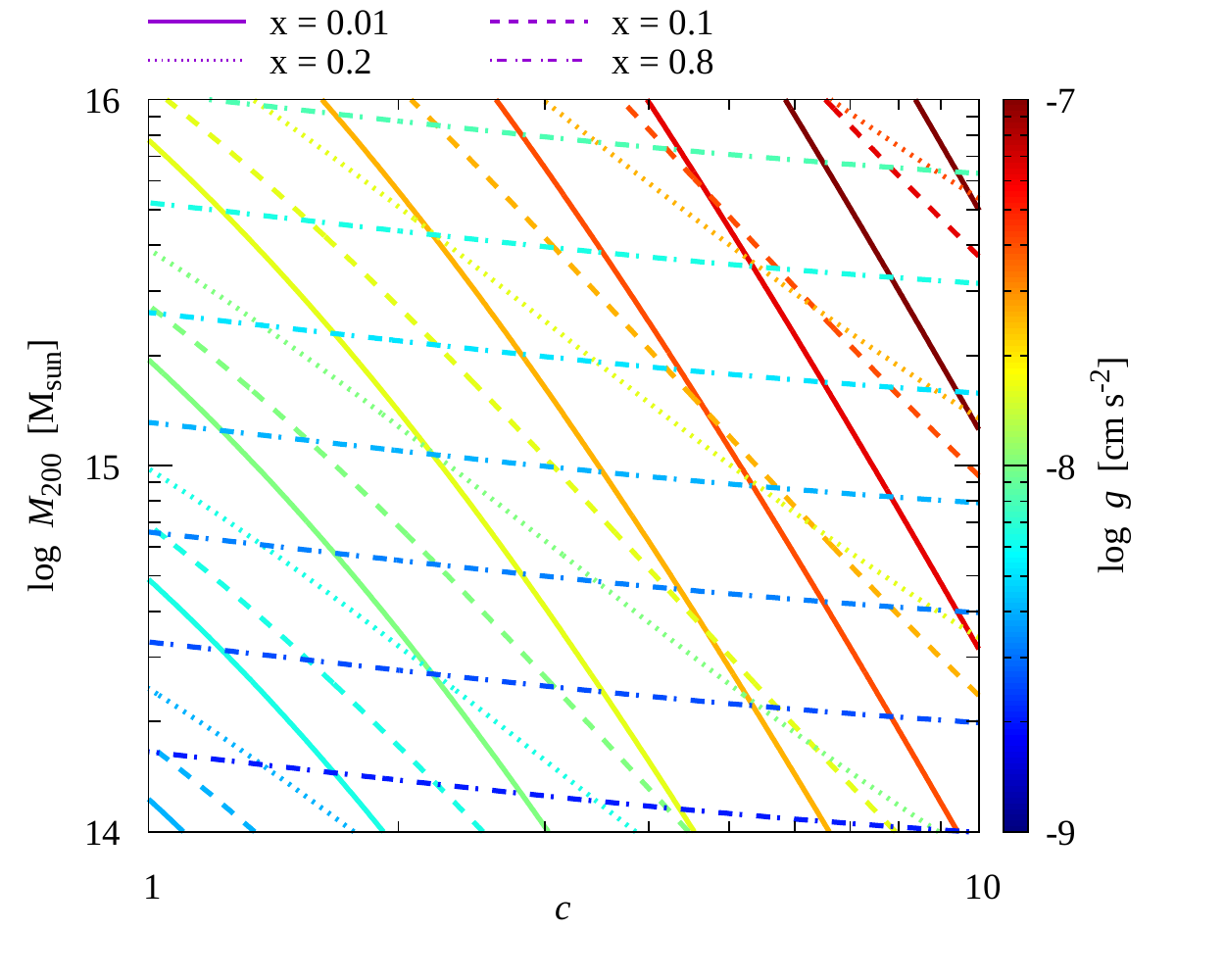}
	\caption{$M_{200}-c$ degeneracy in the NFW profile (\ref{eq:g-nfw}). Different line styles represent different values of $x = r/r_{200}$ and the color of the lines represent the gravitational acceleration, $g$. Although the degeneracy follows Eq. \ref{eq:g-for-small-x} for $x \ll 1$, it remains a concern even for slightly higher $x$. The mass uncertainty becomes negligible at $x\sim 1$.}
	\label{fig:Mv-c-large-x}
\end{figure}

\subsection{Recovering the mass profile}
\label{subsec:recovering-mass}
Often the recovered $g(r)$ profiles from the analysis methods are integrated to infer the mass profiles of the galaxy clusters since $M(r) = \frac{r^2 g(r) }{G} $. To understand the bias in the process, we plot the obtained virial mass ($M_{200}$) and the concentration parameter ($c$) from the \mcfit for all the toy models and the simulated cases in Fig. \ref{fig:M200-c}. The blue points represent all the toy clusters and the orange squares represent the values obtained from the analysis of the simulated data. The hashed blue circle represents the original value used to construct the hydrostatic equilibrium of the toy models, $M_{200} = 2.5 \times 10^{14} M_\odot$ and $c = 4.7$. The red empty circle represents 
$M_{200}$ for the simulations of \citep{prasad18}.  We find  that the recovered values can be off from the actual values almost by a factor of $2-3$ in $c$ and by a factor of $5-10$ in $M_{200}$. This also indicates the uncertainty to which we should believe the cluster mass estimation from the spectral observations of the clusters not going close to the virial radius (both our toy models and simulations extend out to 500 kpc). We stress that, this is due to the systematic uncertainty in the method itself and is very different from the instrumental uncertainties (precision) estimated for the parameters (shown by the error bars against each point). Moreover, there is a systematic  
degeneracy between the virial mass and the concentration parameter, \tt{i.e.,}  $M_{200} c^2 \sim$ const.  This degeneracy can be anticipated from Eq. \ref{eq:g-nfw} for $x \ll 1$ or equivalently, $c x \ll 1$ and considering that $r_{200} \propto M_{200}^{1/3}$. 
\begin{eqnarray}
\label{eq:g-for-small-x}
g(x) &\propto & \frac{M_{200}^{1/3}c }{f(c)} \times \frac{1}{x} \left[ \frac{\log(1+cx) }{cx} - \frac{1}{1+c x} \right]\nonumber \\
 &\propto & M_{200}^{1/3}\: c
\end{eqnarray}
where, in the last step we have used $f(c) \propto c$ (roughly valid for $1 < c < 10$). This relation is very close to the obtained degeneracy. It is therefore clear that for X-ray data only going out to small $x = r/r_{200}$, such a degeneracy is natural to show up in the mass estimates.

Although, the above estimate is for $x \ll 1$, the clusters considered here have virial radii of $\approx 1300$ kpc and $\approx 1900$ kpc for the toy models and the simulations, respectively. This means that for toy models, $x < 0.4$ and for the simulated cluster, $x < 0.25$ since we use a box size of $500$ kpc. In any case, we find that the above-mentioned degeneracy remains for $x \lesssim 0.2$ but becomes increasingly flatter for higher $x$. Figure \ref{fig:Mv-c-large-x} shows the $M_{\rm 200}-c$ degeneracy in the NFW potential for different values of the extent of the X-ray data ($x$; corresponding to different line-styles) and acceleration (indicated by the color of the lines). We see that the degeneracy changes from $M_{\rm 200}\: c^{2.2} \sim$ const at $x=0.1$ to $M_{\rm 200}\: c^{1.7} \sim$ const at $x=0.2$ (comparable with fig \ref{fig:M200-c}), and finally vanishes at $x \sim 0.8$. We, therefore, suggest caution while interpreting the total mass obtained from the X-ray spectral fitting methods presented in this paper, if there are not sufficient photons received from close to the virial radius.

\section{Discussion}
\label{sec:discussion}

\subsection{Effect of central resolution}
\label{sec:resolution}
One of the limitations of the observations is the radial bin size over which the spectra are measured. The sizes of the bins are often determined by considering a minimum X-ray count in a bin so that there are sufficient photon statistics for estimating the density and temperature. Given that the min(\tctf) often arises within the central $\sim 20$ kpc, not resolving this region creates a problem specifically for distant clusters. It limits our ability to look into the very central region where \tc may be significantly different  
from the value at larger radii. To investigate this possibility, we re-do our fits for the \textit{simt840} case but with a larger central bin-size. We increase the central bin size from $2.5$ kpc (default; case R1 in fig \ref{fig:tctf_comp}) to $5.5$ kpc (case R2) and $10$ kpc (case R3). We find that the \tctf increases only by a factor of 2 at most (see right panel of fig \ref{fig:tctf_comp}). Given that case R1 is able to recover a min(\tctf) value of close to 2-3, we conclude that coarsening the radial bins would not immediately move the cluster above the \tctf$=10$ line. We also notice in the right panel of fig \ref{fig:tctf_comp} that coarsening the radial resolution has little effect in a state with $5\lesssim$ \tctf $\lesssim 10$. In other words, a \tctf $\lesssim 10$ cluster should always be detected as a \tctf$\lesssim 10$ cluster. 

\subsection{Observational energy band}
\label{sec:softXray}
Throughout our paper, we  have used spectra with a bandwidth of $0.1-15$ keV, whereas, most of the analysis of the Chandra spectra is done for the energy range of $0.7-12$ keV. Ideally, such observational bias can neglect the contribution of the $T \lesssim 5\times 10^6$ K gas towards overall cooling in that region. Since cooling increases with decreasing temperature in $\sim 10^5-5\times 10^6$ K range (especially in isobaric conditions), missing soft-X-ray photons can underestimate cooling by the lower temperature gas and thus overestimate the \tctf ratio. 

We perform a few experiments on the simulation data to understand the effect of the observational energy band. We trim the spectra for $t = 0.84, 1.6$, and $1.8$ Gyr (corresponding min(\tctf) values are $\approx 3, 12$, and $6$) to $0.7-12.0$ keV range \footnote{Although Chandra energy band is $\approx 0.7-7$ keV, usually one does not have many photons past 7 keV. Therefore, we do not expect any significant change in the \xspec fits even though we included slightly higher energy bins.} and then perform the spectral fitting procedures as described in section \ref{sec:method}. We also use a coarse spatial binning to see the  
combined effect on the \tctf values. The result is shown in the right panel of fig \ref{fig:tctf_comp}. It shows that while higher min(\tctf) values do not change much due to a coarser spatial binning or a harder spectral energy band, the lower min(\tctf) values can be overestimated  
at most by a factor of $2-3$. Our experiments, therefore, suggest that observations in softer x-ray emission ($\lesssim 1$ keV)
are required to recover the min(\tctf)$\lesssim 5$ clusters. However, Chandra observations at such low energies are difficult due to a) the contamination that has built on the Chandra optical blocking filters, reducing the effective area for energies less than $1$ keV, 
and b) additional absorption by the foreground.

\begin{figure*}
    \centering
    \includegraphics[width=\textwidth]{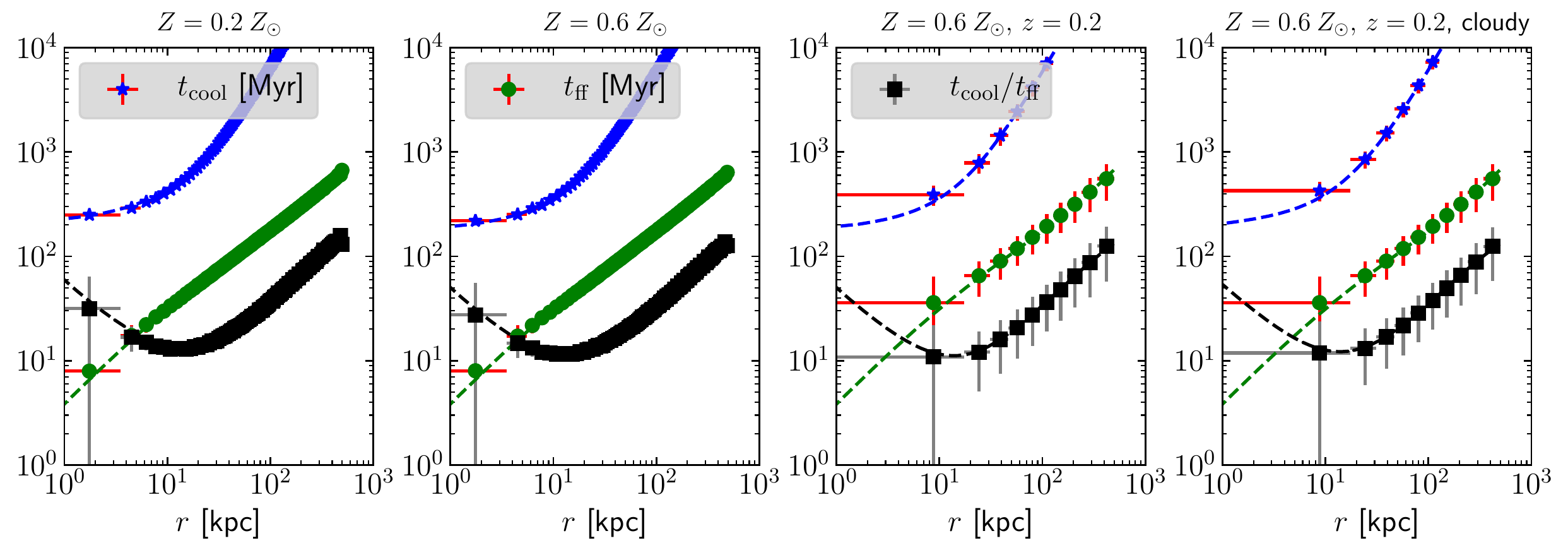}
    \includegraphics[width=\textwidth]{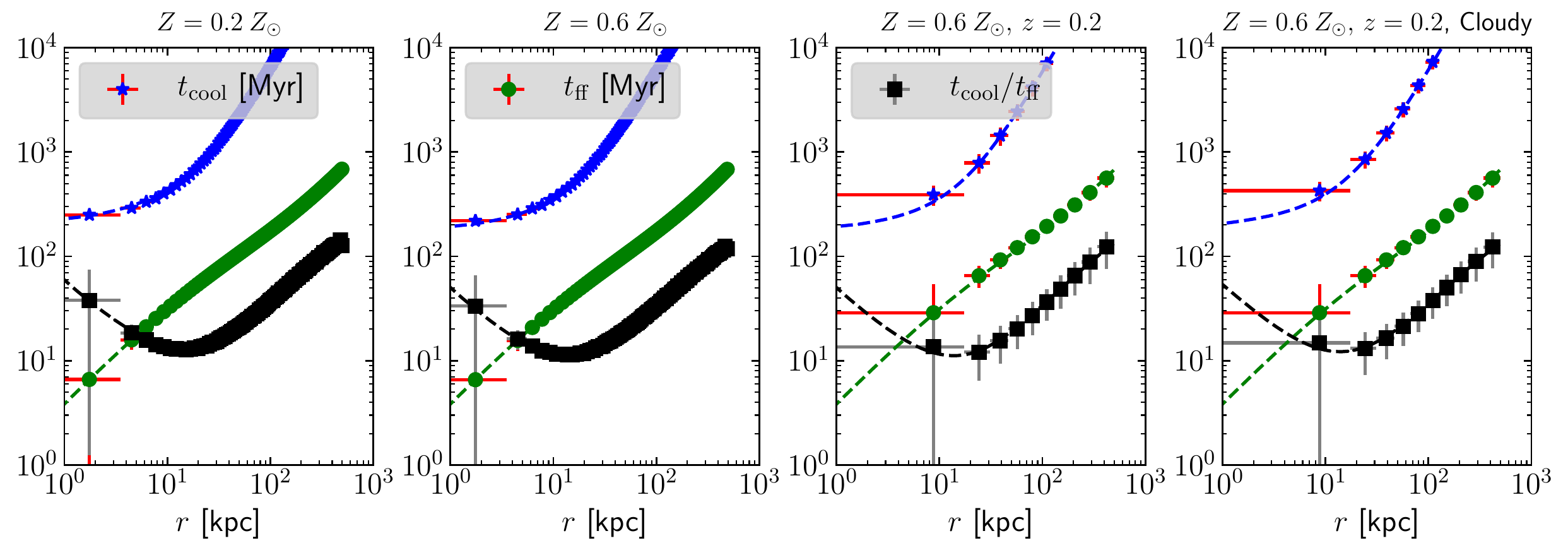}
    \caption{Effects of including realistic photon statistics and redshift in recovering \tc, \tf, and \tctf. The right column also shows the results if we consider the \textsc{cloudy}-17 cooling rates. \textit{Top panels} show the results for the \nlfit method and the \textit{bottom panels} show the results for the \mcfit method. The dashed lines show the actual profiles for the fiducial toy model.}
    \label{fig:error-NL}
\end{figure*}

\subsection{Photon statistics and error in real observations}
\label{sec:photonstats}
So far, we have not considered realistic photon statistics while fitting the spectra in \xspec, a problem often faced in real observations due to the low number of x-ray photons. To understand what the photon statistics may look like, we perform some tests on our fiducial toy model (as presented in section \ref{subsec:method-testing}) by assuming that the photons are detected through an instrument with an effective area, $A_{\rm eff} = 100$ cm$^2$, exposure time $t_{\rm exp} = 300$ ksec, and that the photons suffer through Poisson statistics. The statistics is included by switching on \textit{cstat} while fitting in  \xspec. The analysis is repeated by assuming a higher metallicity ($Z = 0.6 Z_\odot$) as well as a realistic redshift ($z = 0.2$) for the cluster. In the high redshift case, we binned the photons in coarser spatial and energy bins due to the low photon count. The results of this experiment are shown in figure \ref{fig:error-NL}. In the \nlfit method (upper panel), the introduction of \cstat does not alter the results for the low redshift cases. Additionally, the error bars are also negligible (owing to a large photon count), except at the innermost radius where the error in \tf is slightly higher due to a large cell width to radius ratio. For the high redshift cases, the photon statistics become important even after we bin the data in coarser spatial and energy bins. Fitting the spectra with a low photon number, thus introduces a non-negligible error in both \tc and \tf, which is then propagated to \tctf (see sec \ref{appsec:error-calc} for details of the error calculation). The error bars, however, are mostly negligible in \tc and \tf, except at the radius where \tctf error can be close to 30\%. Overall, The recovered values also follow the actual values quite well.

The uncertainties obtained from \xspec toward the density and temperature are also used in the \mcfit method to obtain fits for $\rho_0$, $M_{200}$, and $c$. The introduction of the photon statistics does not significantly improve/worsen the fit for the above quantities, as can be seen from the mean values recovered for \tc, \tf, and \tctf (bottom panel of fig \ref{fig:error-NL}). The error in $\rho_a$, $M_{200}$, and $c$ are also comparable to the fits without the photon statistics ($\sim 10-20\%$). The recovered profiles overall track the actual profiles in the toy model. The error bars in the recovered quantities are also $\lesssim 30\%$ except for the innermost radius where the spatial resolution is the main source of error in estimating \tf and \tctf.

\subsection{Discrepancy between observations and simulations}
\label{subsec:obs-sim-disc}
The biases introduced by observational/modeling limitations (limited angular/spatial resolution, lack of soft X-ray spectra, assumed the functional form of the potential, etc.) tend to overestimate \tctf by a factor of $\lesssim 3$ in the worst-case scenario that we explore, especially for the smallest \tctf. 

Figure 11 in \citet{prasad18} shows a comparison of the distribution of simulated (with and without BCG potential and star formation rate $\sim 5 M_\odot~{\rm yr}^{-1}$) and observed (reported in \citealt{pulido18}) cool core clusters. The observational sample from \citet{pulido18} shows 55 clusters and groups with molecular mass between $10^8$ and $10^{11} M_\odot$, and star formation rate between 0.5 and 270 $M_\odot~{\rm yr}^{-1}$. In contrast to simulations, the observed histogram shows a clear lack of systems with \mtctf$\lesssim 5$. This may be largely due to the overestimation of \tctf in the core because of insufficient spatial resolution (section \ref{sec:resolution}) and the lack of soft X-ray spectrum (section \ref{sec:softXray}) which mostly affect the coolest cores. The same figure in \citet{prasad18} shows that the peaks of the observed and simulated \mtctf histograms also do not match. The observational peak is narrower and occurs at a slightly higher value (\mtctf$\approx 11$) compared to the simulations (NFW potential run peaks at \mtctf$\approx 9$ and the NFW+BCG run has a broad peak between 7 and 11). Thus the observed and simulated X-ray properties of cool cores can be reconciled.

The observational samples of cool core clusters and groups have different selection criteria (e.g., \citealt{voit2015,lakhchaura16,pulido18}), and the distribution of cool-core properties may differ because of that. But the biggest discrepancy from simulations in most of them is the lack of systems with \mtctf$\lesssim 5$, which can be reconciled as discussed in the previous paragraph.

A related recent puzzle is the discovery of a cooling flow in the core of the Phoenix cluster with \mtctf$\approx 1$ in its very center (\citealt{mcdonald19}). How can we explain such a system when observations indicate a general absence of systems having cool cores with \mtctf$\lesssim 5$? Phoenix, despite being a high redshift cluster ($z \approx 0.6$), has been observed with an exquisite spatial resolution (it was possible to use two radial bins within 10 kpc; this allowed \citealt{mcdonald19} to properly separate the AGN and cluster core emission). Moreover, it is a massive/hot cluster with a peak temperature of $\approx 14$ keV. Being hot, the determination of density and temperature in the dense/cool core is not affected as much by the absence of soft X-ray spectra.

Last but not the least, the simulations still very crudely model (or even totally ignore) very important physical processes such as thermal conduction, magnetic fields, angular momentum transport within the central $\sim$ kpc (\citealt{gaspari13,prasad17}), which can affect the distribution of \mtctf. Simulations may not only be missing important physics, but they also have not exhaustively explored various important parameters (such as feedback efficiency).

\section{Conclusions}
\label{sec:conclusions}
In this paper, we have studied various models of the intracluster medium using some standard techniques of X-ray spectral analysis. We have created toy models resembling different kinds of non-hydrostatic features such as cavities, inhomogeneities, unknown potential, etc. to create projected X-ray spectra. We then deproject them using often-used tools \citep[especially,][]{Nulsen2010, lakhchaura16} to recover various  
physically important quantities such as \tc, \tf and \tctf. Comparing these recovered quantities with the values from the input models allow us to examine the bias introduced by X-ray observations and the analysis methods. We also use our projection-deprojection method on realistic simulation data to obtain min(\tctf) during various  
stages of AGN activity and compare them with the ratio obtained from observational samples \citep[such as][]{hogan17, pulido18}. Below we summarize the main findings of our paper:

\begin{itemize}
\item  The non-linear fitting method, used by \citet{lakhchaura16} and the MCMC method (described by \citealt{Nulsen2010}) used by more recent studies (such as \citealt{hogan17}) recover the min(\tctf) of the non-cooling as well as the cool-core clusters fairly well. The presence of large cavities leads to overestimating the \tc (and hence \tctf) in both methods. The amount of overestimation depends on the solid angle subtended by the cavity at the cluster center. The \nlfit method works better for the systems where the gravitational potential at the center is not known accurately. It also better estimates the cooling time.

\item We are able to recover min(\tctf) as low as $2-3$ by analyzing the simulation data. We, however, find a factor of $\lesssim 3$ overestimation of the min(\tctf) if the spatial/angular resolution is $\gtrsim 10$ kpc and if soft X-ray spectrum is not available. 
The observed lack of  
cool-core clusters below min(\tctf) $< 5$ \citep{pulido18} compared to the  
value from AGN simulations may arise due to poor spatial resolution and the lack of soft X-ray spectra. Additionally, there may be additional heating  
in cool-core clusters that can raise the \tctf value above that seen in AGN feedback simulations.

\item The estimated cluster mass (M$_{200}$) from the \mcfit method suffers from a systematic degeneracy with the estimated concentration parameter, $c$. Such degeneracy can cause the dark matter mass estimates to be off by a factor of $5-10$ from the original value and arises purely due to the form of the NFW profile at small $r/r_{\rm 200}$ ($\lesssim 0.2$) values. We stress that this degeneracy goes away if one considers the X-ray data for $r/r_{\rm 200} \sim 1$ (as used in \cite{nagai2007}). This also emphasizes the need for X-ray observations from close to the virial radius of a galaxy cluster in order for them to be used as useful cosmological probes, which requires an accurate cluster mass. 
\end{itemize}

This paper, therefore, dives deep into understanding the biases of the analysis methods often used to study the intracluster medium. Although our simplistic toy models may not cover all the possible complications arising in a real ICM, it provides an overall idea of how much various complications can affect the estimated \tctf. We hope that the current study will be a useful benchmark to characterize such biases.

\section*{Acknowledgment}
We thank Nazma Islam for her help in solving technical issues in \xspec. We also thank the anonymous referee for the helpful comments on the paper.
KCS acknowledges support from the Israeli Science Foundation (ISF grant no. 2190/20). PS acknowledges a Swarnajayanti Fellowship (DST/SJF/PSA-03/2016-17) and a National Supercomputing Mission (NSM) grant from the Department of Science and Technology, India.

\section*{Data availability}
All the data/codes used in this paper can be found in this paper or under their cited references.

\bibliographystyle{mnras}
\bibliography{bibtex}

\appendix
\section{Error calculation}
\label{appsec:error-calc}
Errors in \tc, \tf, and in \tctf ratio are determined numerically from the obtained uncertainties in the fit parameters 
\textit{viz.} $P_0,\: a,\: \alpha_1,\: \alpha_2,\: T$, and $n_e$ (for \nlfit) and $M_{200},\: c,\: T$, and $n_e$ (for \mcfit). The numerical uncertainty for any quantity $Q \equiv Q(r, p_1, p_2, ...)$ is given as
\begin{equation}
    \Delta Q = \frac{\partial Q}{\partial r} \Delta r 
      + \frac{\partial Q}{\partial p_1} \Delta p_1 
      + \frac{\partial Q}{\partial p_2} \Delta p_2 + ...
\end{equation}
where, $p_1,\: p_2,\: ...$ are the independent parameters and $\Delta p_1, \: \Delta p_2,\: ...$ are the corresponding uncertainties. The derivatives have been calculated numerically.

\begin{figure}
    \centering
    \includegraphics[width=0.5\textwidth]{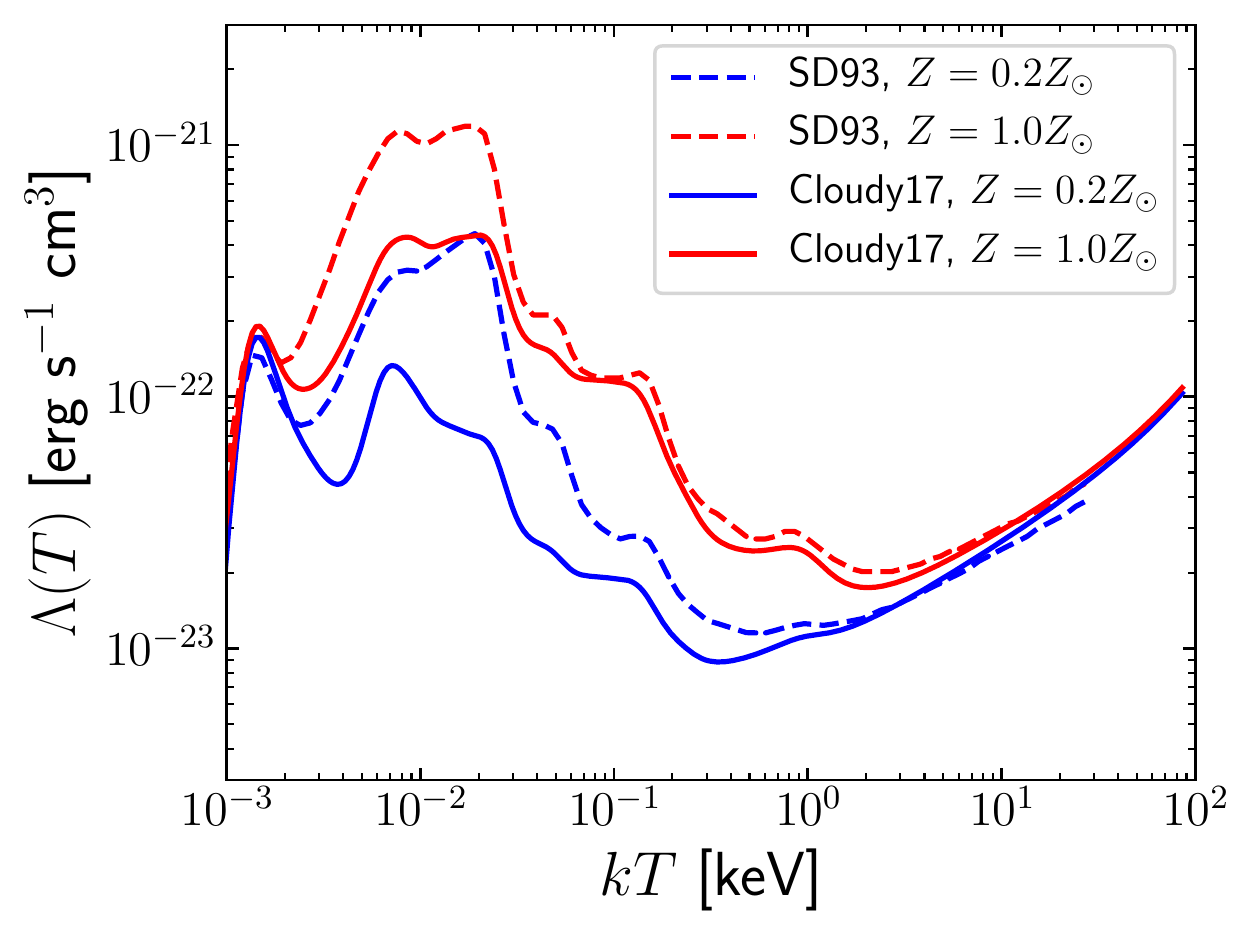}
    \caption{Comparison of the cooling functions of \citealt{sutherland+dopita+93} (dashed lines) and a more recent cooling curve from \textsc{cloudy}-17 \citep[solid lines][]{cloudy2017}. Different colors represent different metallicities. The discrepancy of cooling rates above $\gtrsim 0.5$ keV is almost negligible.}
    \label{appfig:cooling-comparison}
\end{figure}

\label{lastpage}
\end{document}